\documentclass{article}
\usepackage{fancyhdr}

\usepackage{setspace}
\usepackage{fullpage} 
\usepackage{multirow} 
  \usepackage{natbib}
  \bibpunct{(}{)}{;}{a}{}{,}

\usepackage{amsmath}
\usepackage{amsfonts}
\usepackage{graphicx}
\usepackage{subfigure}
\usepackage{pst-all}

\usepackage{verbatim}
\usepackage[mathscr]{eucal}

\newcommand{\argmax}{\ensuremath{\arg\max}}

\newcommand{\mM}{\ensuremath{\mathcal{M}}}
\newcommand{\sm}{\ensuremath{\setminus}}

\newcommand{\candidateset}{\ensuremath{V \sm \mM'}} 
\newcommand{\redcandidates}{\ensuremath{\mM \sm \mM'}} 
\newcommand{\greenvertices}{\ensuremath{V \sm \mM}} 

\begin{document}

\begin{centering}
\Large{Vertex Nomination via Content and Context}

\vspace{20pt}
\today

\vspace{20pt}
\large{Glen A. Coppersmith and Carey E. Priebe}

Human Language Technology Center of Excellence

Johns Hopkins University

\vspace{20pt}

\end{centering}

\section*{Abstract}

If I know of a few persons of interest, how can a combination of human language technology and graph theory help me find other people similarly interesting? If I know of a few people committing a crime, how can I determine their co-conspirators? Given a set of actors deemed interesting, we seek other actors who are similarly interesting. We use a collection of communications encoded as an attributed graph, where vertices represent actors and edges connect pairs of actors that communicate. Attached to each edge is the set of documents wherein that pair of actors communicate, providing content in context -- the communication topic in the context of who communicates with whom. In these documents, our identified interesting actors communicate amongst each other and with other actors whose interestingness is unknown. Our objective is to nominate the most likely interesting vertex from all vertices with unknown interestingness. As an illustrative example, the Enron email corpus consists of communications between actors, some of which are allegedly committing fraud. Some of their fraudulent activity is captured in emails, along with many innocuous emails (both between the fraudsters and between the other employees of Enron); we are given the identities of a few fraudster vertices and asked to nominate other vertices in the graph as likely representing other actors committing fraud. Foundational theory and initial experimental results indicate that approaching this task with a joint model of content and context improves the performance (as measured by standard information retrieval measures) over either content or context alone.

\break

\section{Introduction}

Given a set of documents containing communications among a collection of actors
and an identified subset of actors deemed interesting, we wish to select
actors from outside the identified set who exhibit similar behavior to the identified interesting actors.
For a concrete example, within the Enron email collection
(see, e.g., \cite{Priebe:2005})
is a set of executives and traders allegedly committing fraud.
If we know the identities of a subset of the fraudsters, can we
nominate other people from the company as likely fraudsters?
We assume that what indicates that actors are interesting (fraudulent) is manifest both
in the topics about which they communicate (the content of their messages) and
with whom in the company they communicate (the context of their messages).
We conceptualize this as an attributed graph, where each vertex is an actor and pairs of actors
that communicate are connected by edges. The edges are attributed by the content of the messages 
exchanged (in our case, represented as a distribution over topics).
We design and evaluate a family of test statistics
that score each actor (vertex) based on the
content and context of their email communications (edges).
We nominate vertices from outside the identified set
as likely to be interesting.
This task has noted similarities to the Netflix challenge (e.g., \cite{BellKor08}),
recommender systems (\cite{ResnickVarian1997} and contents of the special issue), 
and detecting communities of interest (e.g. \cite{CortesPregibonVolinsky2002}).

Information useful for the vertex nomination task might be encoded in both content
and context. It is reasonable to assume
that test statistics based on either content alone or context alone would have
some efficacy for vertex nomination, but statistics which take advantage
of both content and context might provide superior inferential capability (e.g. \cite{PPMCGG}).
Selecting test statistics useful for this task
(or selecting the uniformly most powerful test statistic against some specified composite alternative)
 is both interesting and
decidedly nontrivial.
 (See \cite{PCR2010} for a summary of inferential complexity
 in a related task in perhaps the simplest possible model -- without content.)
We set up a deceptively simple generative model
(described in Section \ref{the_model} and depicted in Figure \ref{vn_kidney_and_egg})
 to study this task and
present results from simulations and experiments on real data (the Enron email corpus).

The possible space of test statistics is practically limitless,
even for this simple setting.
For tractability, we limit ourselves to 
a simple family of linear fusion statistics (Section \ref{simple_statistics}) 
and demonstrate how their performance for this task
depends on many underlying factors
(manifest as parameters in the generative model and
latent qualities of the data).
The optimal performance is found in a fusion of content and context
rather than either alone, in both simulated and observed data
(Sections \ref{simulations} and \ref{observed_experiments} respectively).

This paper proceeds as follows: 
Section \ref{the_model} spells out our assumptions and describes the joint model of content and context,
Section \ref{methods} describes the experimental and evaluation methods used,
Section \ref{simulations} describes simulation experiments where the content and context 
is generated according to our model, 
Section \ref{observed_experiments} demonstrates that (A) our assumptions are reasonable (and real data 
corresponding to the assumptions does naturally occur), (B) when our assumptions are met, vertex nomination works,  
and (C) when our assumptions are met, the fusion of content and context is superior to either alone,
and Section \ref{conclusions} makes concluding remarks and discusses future directions.

The appendix 
details our data set, the Enron email corpus.

\section{Model}
\label{the_model}

We base our model upon two assumptions, detailed below. 
Specifically, when the physical world exhibits a group of interest that 
meets these assumptions, our model is reasonable (as demonstrated in Section \ref{observed_experiments}).
We observe communications among our identified interesting set,
  among our candidate set, and
between actors in the identified set and actors in the candidate set.

\begin{itemize}
\item \textbf{Assumption 1}: Pairs of vertices in the group of interest (identified and not identified)
communicate among themselves with a different frequency
than other pairs.

\item \textbf{Assumption 2}: The group of interest communicates about topics in different proportions than the population of actors as a whole.
\end{itemize}

The context information available for vertex nomination is derived from \textbf{Assumption 1},
while the content information is derived from \textbf{Assumption 2}.

Let $G = (V, E, \phi_V, \phi_E)$ be the simplest of attributed graphs
($G$ is undirected, with no self-loops, no multi-edges and no hyper-edges).
Let $V$ be the set of vertices (actors) and $E$ be the set of edges (communication 
between pairs of actors). Specifically, $E\subset V^{(2)}$,
where $V^{(2)}$ denotes the set of 
unordered pairs of vertices.
Attribution functions $\phi_V: V \to \Phi_V$ and $\phi_E: V^{(2)} \to \Phi_E$ place
(categorical) attributes on the vertices and edges, respectively, where
$\Phi_V = \{1,...,K_V\}$, $\Phi_E = \{0,1,...,K_E\}$ and
$K_V$ is the number of vertex attributes (interesting and not interesting for our purposes) and
$K_E$ is the number of edge attributes (topics for our purposes); 
$\phi_E = 0$ represents a non-observed edge, so for all $e \notin E$, $\phi_E(e)=0$ and for all $e \in E$, $\phi_E(e) \in \{1,...,K_E\}$. 
For this investigation, $K_V=K_E=2$ and $\Phi_V = \Phi_E = \{red, green\}$. We use $red$ and $1$ interchangeably, as appropriate for the context. Likewise for $green$ and $2$.

\begin{figure}
\begin {centering}

\begin{tabular}{|c|}
\hline
\includegraphics[width=0.4\textwidth]{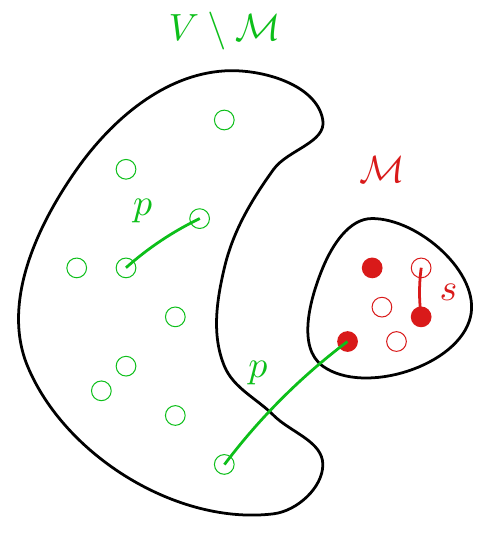} \\
\hline
\includegraphics[width=0.65\textwidth]{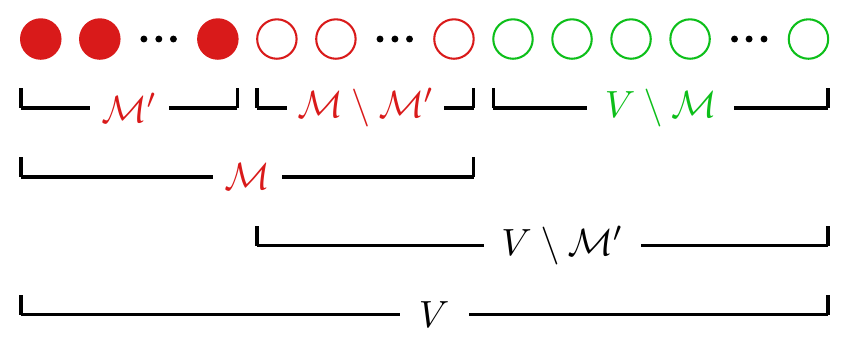} \\
\hline
\end{tabular}

\end{centering}
\caption{\label{vn_kidney_and_egg}
Our model, a $\kappa(n,p,m,s;m')$ graph with $n=|V|$ vertices,
$m=|\mM|$ of which have attribute $red$ and
$n-m = |\greenvertices|$ of which have attribute $green$.
We observe attributes for only $m'=|\mM'|$ {\em identified set} vertices (filled circles) with
the remaining $n-m'=|\candidateset|$ {\em candidate set} vertices (open circles) having occluded attributes.
Edges with attribute $green$ are of the topic not of interest ($2$) and $red$ edges are topic of interest ($1$).
Pairs of $red$ vertices (regardless of occlusion) are connected according to probability distribution over topics $s$,
while pairs of vertices where at least one is labeled $green$ are connected according to $p$.
The vertex nomination task is to select one vertex from the candidate set \candidateset ~
(the vertices with occluded attributes, shown here as open circles)
that is in \redcandidates ~ (truly $red$).}
\end{figure}

For our investigation, we use a simple edge- and vertex-attributed independent edge model.
We use a stochastic block-model random graph (sometimes referred to as a ``kidney-egg'' or $\kappa$ graph), where there
is a ``chatter'' group present -- a subset of the actors which communicate
amongst themselves in excess of what is expected from
the activity present in the rest of the graph and with a topic distribution different from
that governing the rest of the graph.
As depicted in Figure \ref{vn_kidney_and_egg},
$\kappa(n,p,m,s)$ is a random graph model \cite{bollobas:2001} on $n$ vertices ($|V|=n$);
$|\{v:\phi_V(v)=1\}| = m$, so $m$ vertices have the attribute of interest ($red$)
and communicate differently than the collection
$\{v:\phi_V(v)=2\}$ of $n-m$ not of interest ($green$) vertices.
The edge attribute for a pair of vertices $u,v$ with $\phi_V(u)=\phi_V(v)=1$
is governed by the probability vector $s=[s_0,s_1,s_2]'$ where
$s_1$ is the probability that the edge is $red$ ($\phi_E(uv)=1$),
$s_2$ is the probability that the edge is $green$ ($\phi_E(uv)=2$), and
$s_0$ is the probability of no edge;
edge attributes for all other pairs of vertices are governed by $p=[p_0,p_1,p_2]'$.
Like $s$, $p_1$ is the probability of a $red$ edge, 
$p_2$ is the probability of a $green$ edge and 
$p_0$ is the probability of no edge.

The observed graph includes occlusion of most of the vertex attributes:
$G{'} = (V, E, \phi_V, \phi{'}_V, \phi_E)$ is a $\kappa(n,p,m,s;m')$ graph
where
$\phi{'}_V: V \to \Phi_V \cup \{0\}$ and $\phi_V(v)=0$
denotes that the attribute for vertex $v$ is occluded.
For our particular setting, all observed attributes are $red$ and we observe no $green$ attributes
(Figure \ref{vn_kidney_and_egg}).
We let $\mM = \{v:\phi_V(v)=1)\}$ be the set of vertices with true $red$ attributes.
Our {\em identified set} --
the set of vertices with observed (true) $red$ attributes --
is given by $\mM' = \{v:\phi{'}_V(v)=1\}$ and $|\mM'| = m'$.
(We assume that the identified set $\mM' \subset \mM$ is selected at random.)
The candidate set is $V \sm \mM'$. 
We assume that there is no error in the vertex-attributes -- just occlusion;
in addition, we assume that we observe the attributes on all edges, and there is no error in the edge-attributes.

We assume that $n >> m > m' > 0$.
That is, there is at least one vertex known to be
of interest ($m' \geq 1$), which allows the set of context measures
we employ to measure functions of the graph-proximity to a member of $\mM'$.
We also assume that candidate set $V \sm \mM'$ contains
at least one true $red$ ($m>m'$) and at least one true $green$ ($n>m$) vertex.
(The question of whether or not there exist any $red$ vertices in the candidate set
is an interesting one; we do not directly address it here, but the methods described here
do inform how one might approach that question.)

To rephrase the inference task in our freshly minted notation:
We are given a graph $G'$ on vertices $V$,
 $m$ of which have attribute $red$ ($\mM \subset V$) and
 $n-m$ of which have attribute $green$ (\greenvertices).
All vertex-attributes are occluded save $m'$ drawn from the set $\mM$ ($\mM' \subset \mM$); thus all observed vertex-attributes are $red$.
We wish to
rank order all vertices with occluded attributes -- the candidate set {\candidateset} -- according to their similarity to the identified
set $\mM'$.
Performance is judged by how high in the ranked list of candidate vertices {\candidateset}  
the vertices \redcandidates\ with occluded
(but truly $red$) attributes  fall.

\section{Methods}
\label{methods}

\subsection{Statistics}
\label{simple_statistics}

We employ test statistics, based on the content and context of each vertex and its 
communications, to rank-order the candidate set \candidateset\  
for nomination.
Consider a vertex $v$ in the candidate set \candidateset.
If $p_2 = s_2$ and $p_1 < s_1$,
then $v \in \redcandidates$\ will have a stochastically larger value for both
the number of known $red$ vertices adjacent to $v$ and
the number of $red$ edges incident to $v$.
This observation gives rise to the observation that the
posterior probability of class membership
$\rho(v) = P[\phi_V(v)=1|G',\phi'_V(v)=0]$
is monotonically increasing in both
the context-only statistic
\begin{equation}
T^{0}(v)=\displaystyle\sum_{u\in \{w:wv \in E\}} \mathbb{I}\left\{ \phi'_V(u) = 1 \right\}
\end{equation}
and
the content-only statistic
\begin{equation}
T^{1}(v)=\displaystyle\sum_{uv \in E} \mathbb{I}\left\{ \phi_E(uv) = 1 \right\}.
\end{equation}
This in turn motivates the class of {\em linear fusion statistics}
\begin{equation}
T^{\gamma}(v) = (1-\gamma) T^0(v) + \gamma T^1(v).
\end{equation}
Larger scores are more indicative of membership in $\mM$.
The parameter $\gamma$ determines the relative weight of content and context information.
We rank each vertex $v$ in the candidate set \candidateset\ for nomination as a likely
member of $\mM$ according to $T^{\gamma}(v)$. 
Let $\gamma^\star$ denote the fusion parameter which yields the highest performance.

For our independent edge model $\kappa(n,p,m,s;m')$,
the joint distribution of $T^{0}(v),T^{1}(v)$ is available:
for $v \in \greenvertices$, we have
\begin{eqnarray*}
T^0(v;G)  &\sim& Bin(m',p_1+p_2), \\
T^1(v;G)  &\sim& Bin(n-1,p_1), \\
T^1|T^0=c &\sim& Bin(c,\frac{p_1}{p_1+p_2}) +_{ind} Bin(n-1-m',p_1),
\end{eqnarray*}
while for $v \in \redcandidates$, we have
\begin{eqnarray*}
T^0(v;G)  &\sim& Bin(m',s_1+s_2), \\
T^1(v;G)  &\sim& Bin(m-1,s_1) +_{ind} Bin(n-m,p_1), \\
T^1|T^0=c &\sim& Bin(c,\frac{s_1}{s_1+s_2}) +_{ind} Bin(m-1-m',s_1) \\&~&+_{ind} Bin(n-m,p_1).
\end{eqnarray*}

For each candidate $v \in \candidateset$, we calculate $T^{\gamma}$, where $\gamma \in (0,1)$.
For some plots we select a few illustrative values of $\gamma$, rather than plotting the entire range:
$\gamma=0$ for context-only (represented on plots by ``X''),
$\gamma=1$ for content-only (represented on plots by ``N''),
$\gamma=0.5$ for (one particular instantiation of) fusion of content and context (represented on plots by ``+''), and
$\gamma=\gamma^\star$ for the linear fusion of content and context with the optimal performance (represented on plots by ``*'').

For a given $\gamma$, we rank vertices for nomination according to $T^{\gamma}$ and consider the ordered candidates
$v^{\gamma}_{(1)}, v^{\gamma}_{(2)},\cdots, v^{\gamma}_{(n-m')}$.
E.g., considering vertices in the candidate set \candidateset, we have
$v^{\gamma}_{(1)} = \argmax_v T^{\gamma}(v)$,
$v^{\gamma}_{(2)}$ is the vertex associated with the second largest value of $T^{\gamma}(v)$,
etc.
We evaluate the efficacy of each $T^{\gamma}$ according to three evaluation criteria,
described below.

\subsection{Evaluation Criteria}
\label{EC}

\begin{itemize}

\item \textbf{Probability Correct}
\label{1isRed}

If we nominate {\em one} vertex based on the values of the linear fusion statistic,
then performance can be measured based on whether this nominee is in fact truly red -- the success at rank 1 (S@1): 
\begin{equation}
 \mbox{S@1}(\gamma) = \mathbb{I}\{v^{\gamma}_{(1)} \in \redcandidates\}.
\end{equation}
For a random experiment, we consider $E\left[\mbox{S@1}(\gamma)\right]$.


\item \textbf{Mean Reciprocal Rank}
\label{RMinR}

Reciprocal rank (RR) is a measure of how far down a ranked list 
one must go to find the first truly red vertex:
\begin{equation}
\mbox{RR}(\gamma) = \left(\min\{i: v^{\gamma}_{(i)} \in \redcandidates\}\right)^{-1}.
\end{equation}
For a random experiment, we consider the mean reciprocal rank
\begin{equation}
\mbox{MRR}(\gamma)=E\left[\mbox{RR}(\gamma)\right].
\end{equation}

\item \textbf{Mean Average Precision}
\label{MAP}

Average precision (AP) 
examines the placement within a ranked list of all truly red vertices -- 
the average of the precision at the rank of each truly red vertex.
We define precision at rank $r$ as
\begin{equation}
Pre(r,\gamma) =  \frac{\displaystyle\sum_{i=1}^r\mathbb{I}\{v^\gamma_{(i)} \in \redcandidates\}}{r}
\end{equation}
and average precision as
\begin{equation}
\mbox{AP}(\gamma) = \frac{\displaystyle\sum_{i=1}^{|\candidateset|}\mathbb{I}\{v^\gamma_{(i)} \in \redcandidates\}Pre(i,\gamma)}{|\redcandidates|}. 
\end{equation}
For a random experiment, we define the mean average precision 
\begin{equation}
\mbox{MAP}(\gamma) = E\left[\mbox{AP}(\gamma)\right]. 
\end{equation}


\end{itemize}

Section \ref{simulations} demonstrates that these measures are all highly correlated, as is further explored by \cite{BuckleyVoorhees_UseMAP}. 
For our experiments, the relative ranking of vertex nomination methods is consistent across evaluation measures.

\section{Simulation Experiments}
\label{simulations}

We evaluate the performance of content and context fusion via simulation
in the $\kappa(n,p,m,s;m')$ model.

We consider $p,s$ in the standard 2-simplex $S^2=\{x \in \mathbb{R}^3: x_i \geq 0, \sum_i x_i=1\}$,
constrained so that $s_2=p_2$
(so the probability of $green$ content being present is the same throughout the graph)
and $s_1 > p_1$
(so the probability of $red$ content being present is greater for edges that connect pairs in {\mM} than for all other pairs).
(Note that this implies $s_0 < p_0$, so also overall connectivity probability
is greater for edges that connect pairs in {\mM} than for all other pairs.)
We use $n=184$ (the number of actors in our Enron email corpus) and 
consider various values of $m,m'$ such that $n >> m > m' > 0$.
We assess performance using the three evaluation criteria,
$E[\mbox{S@1}(\gamma)]$,
$\mbox{MRR}(\gamma)$, and
$\mbox{MAP}(\gamma)$,
introduced in Section \ref{EC},
for $\gamma \in [0,1]$.

\begin{figure}
\begin{centering}

\includegraphics[width=400pt]{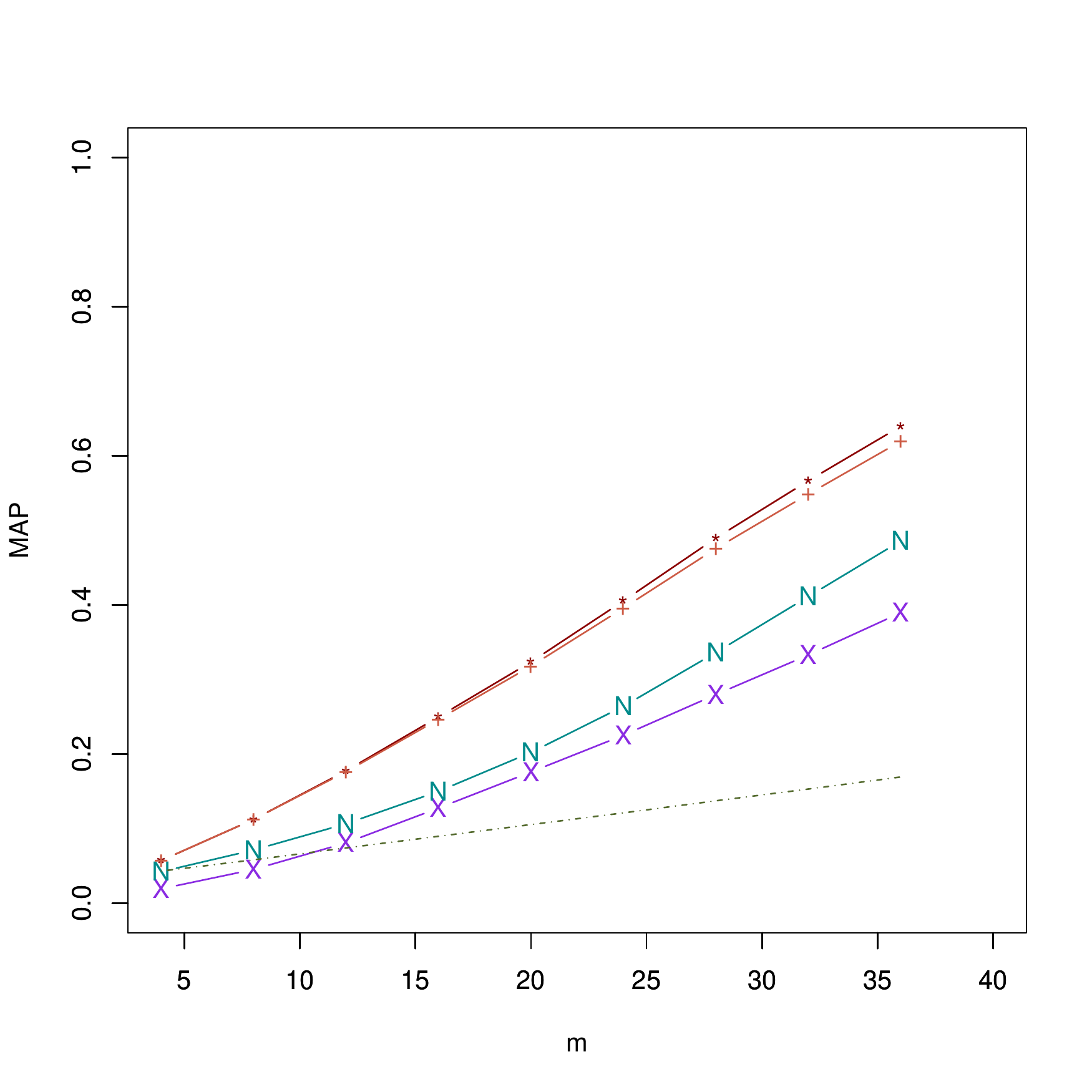}

\end{centering}

\caption{\label{lg_sim_bymetric}
Context, content, arbitrary linear fusion, and optimal linear fusion ($\gamma=\{0,1,0.5,\gamma^\star\}$ respectively)
results for
$\mbox{MAP}(\gamma)$
in the
$\kappa(n=184, p=[0.6,0.2,0.2]', m, s=[0.4,0.4,0.2]'; m'=m/4)$ model, as we vary $m$.
We plot $m$ on the $x$-axis and $\mbox{MAP}(\gamma)$ on the $y$-axis.
Content ($\gamma=1$) is represented by points labeled ``N'',
context $(\gamma=0)$ by points labeled ``X'',
arbitrary linear fusion ($\gamma=0.5$) by points labeled ``+'',
and optimal linear fusion ($\gamma=\gamma^\star$) by points labeled ``*''.
Results are obtained via $1000$ Monte Carlo replicates.
The green dashed line denotes chance performance.}

\end{figure}

Figure \ref{lg_sim_bymetric} presents performance
using $\mbox{MAP}(\gamma)$
for
$$\kappa(n=184, p=[0.6,0.2,0.2]', m, s=[0.4,0.4,0.2]'; m'=m/4)$$ as we vary $m$.
For small $m$, all our fusion statistics perform equally poorly.
(The far leftmost point in
Figure \ref{lg_sim_bymetric}
represents $m=4$ and $m'=1$, where almost no information is available.)
As $m$ (and hence $m' = m/4$) increases,
fusion of content and context provides superior performance:
$\gamma=0.5$ and $\gamma=\gamma^\star$ are superior to either $\gamma=1$ or $\gamma=0$ alone.

\begin{figure}
\begin{centering}

\begin{tabular}{ccc}

$m'= \frac{m}{4}$ &
$m' = \frac{m}{2}$ &
$m' = \frac{3m}{4}$ \\

\includegraphics[width=0.333\textwidth]{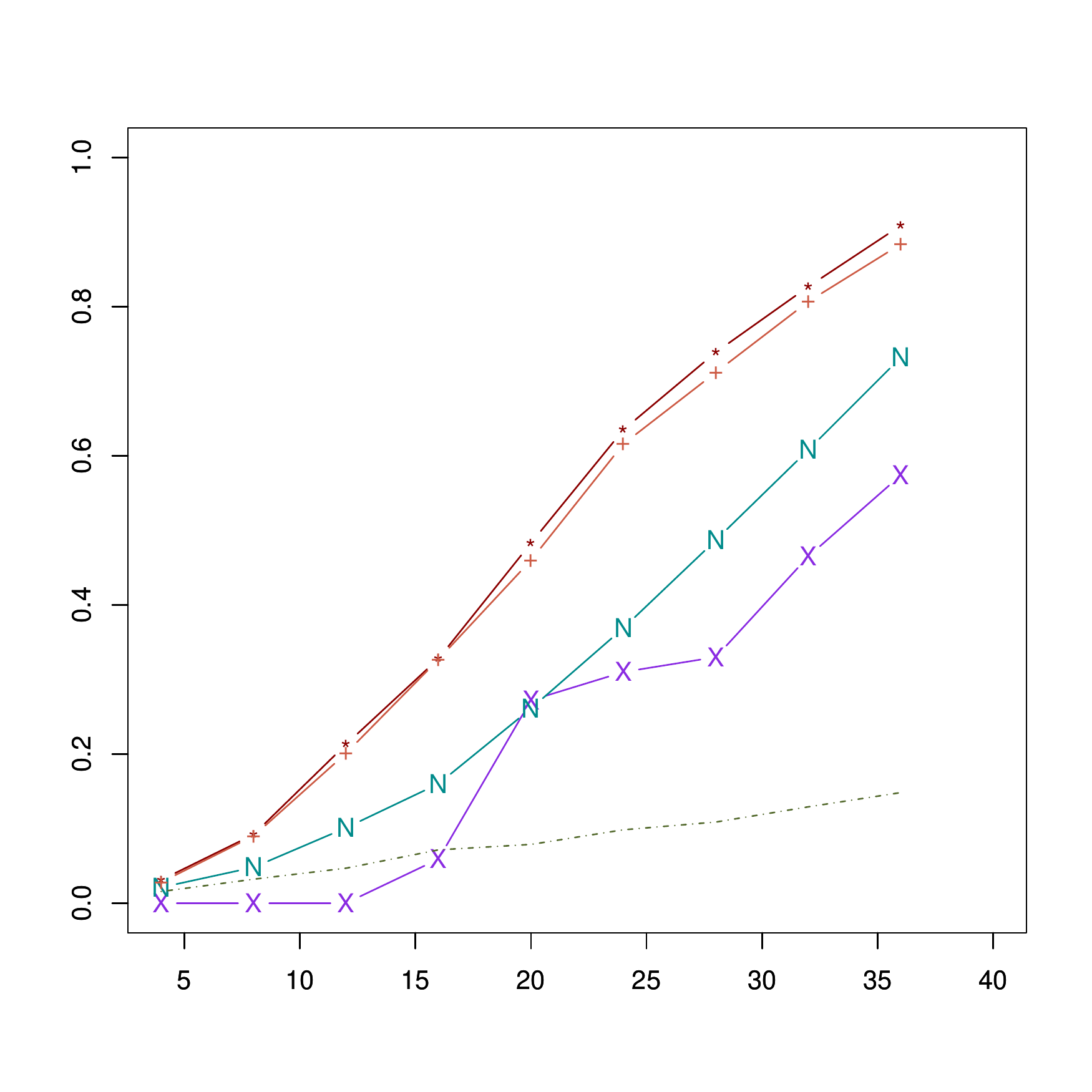} &
\includegraphics[width=0.333\textwidth]{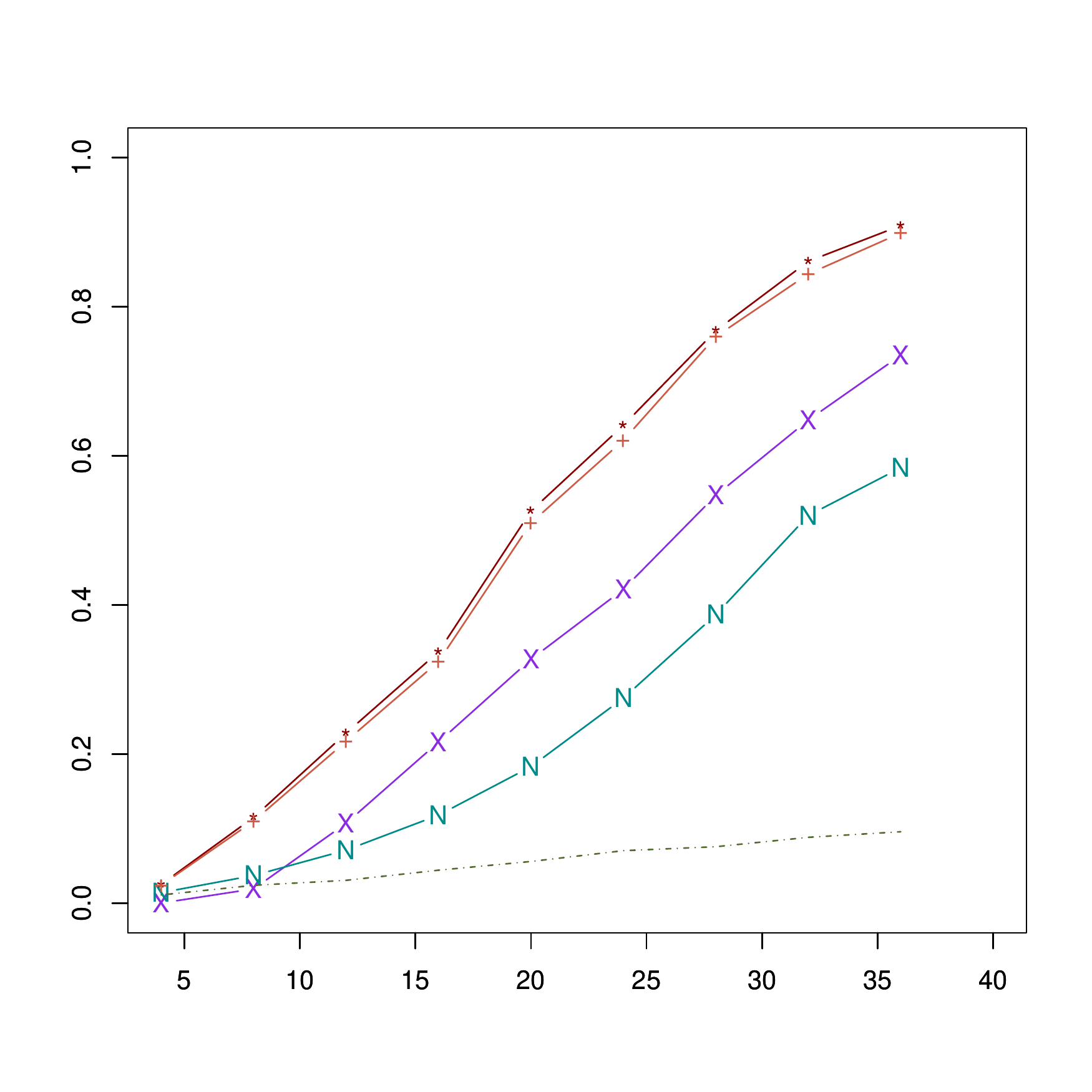} &
\includegraphics[width=0.333\textwidth]{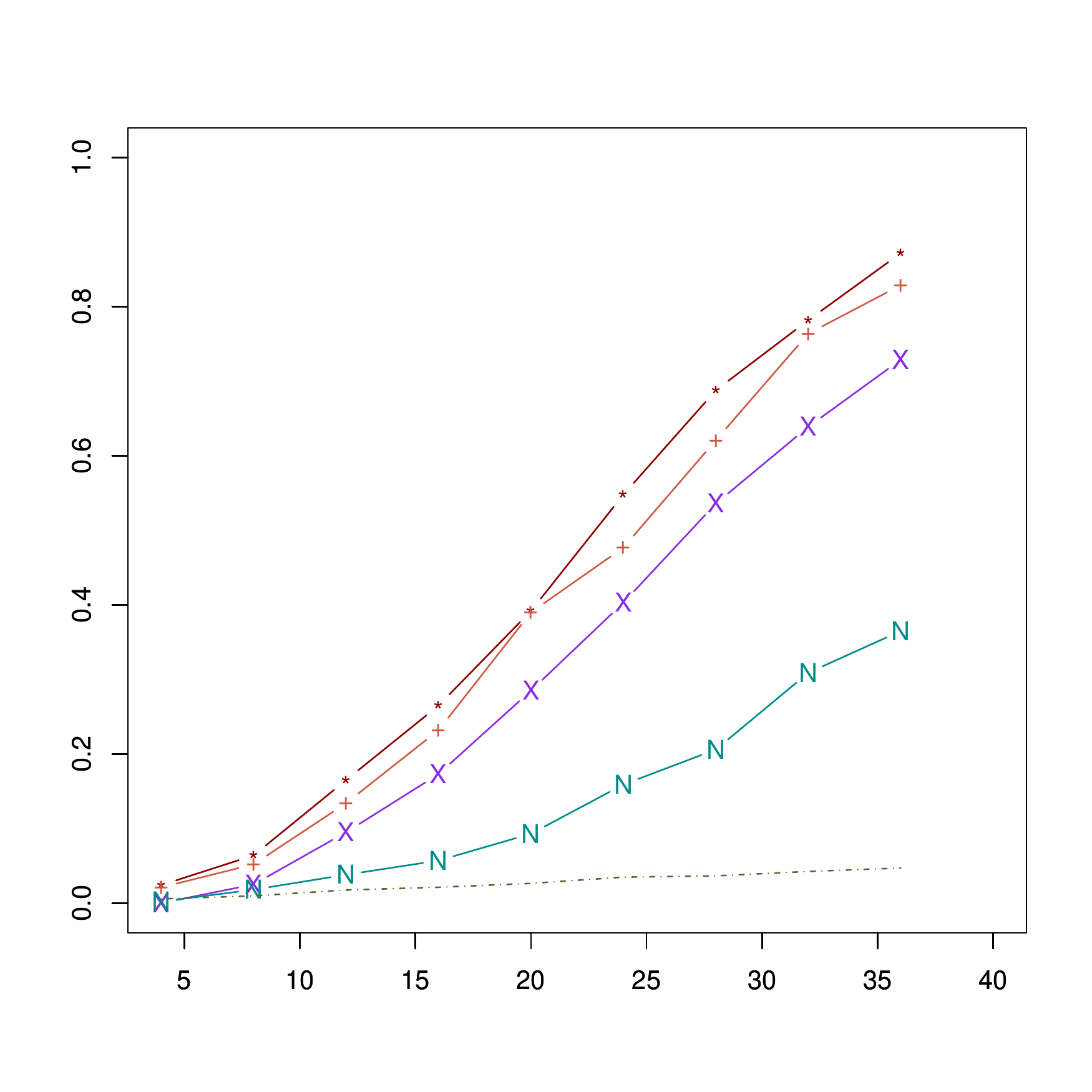} \\

\includegraphics[width=0.333\textwidth]{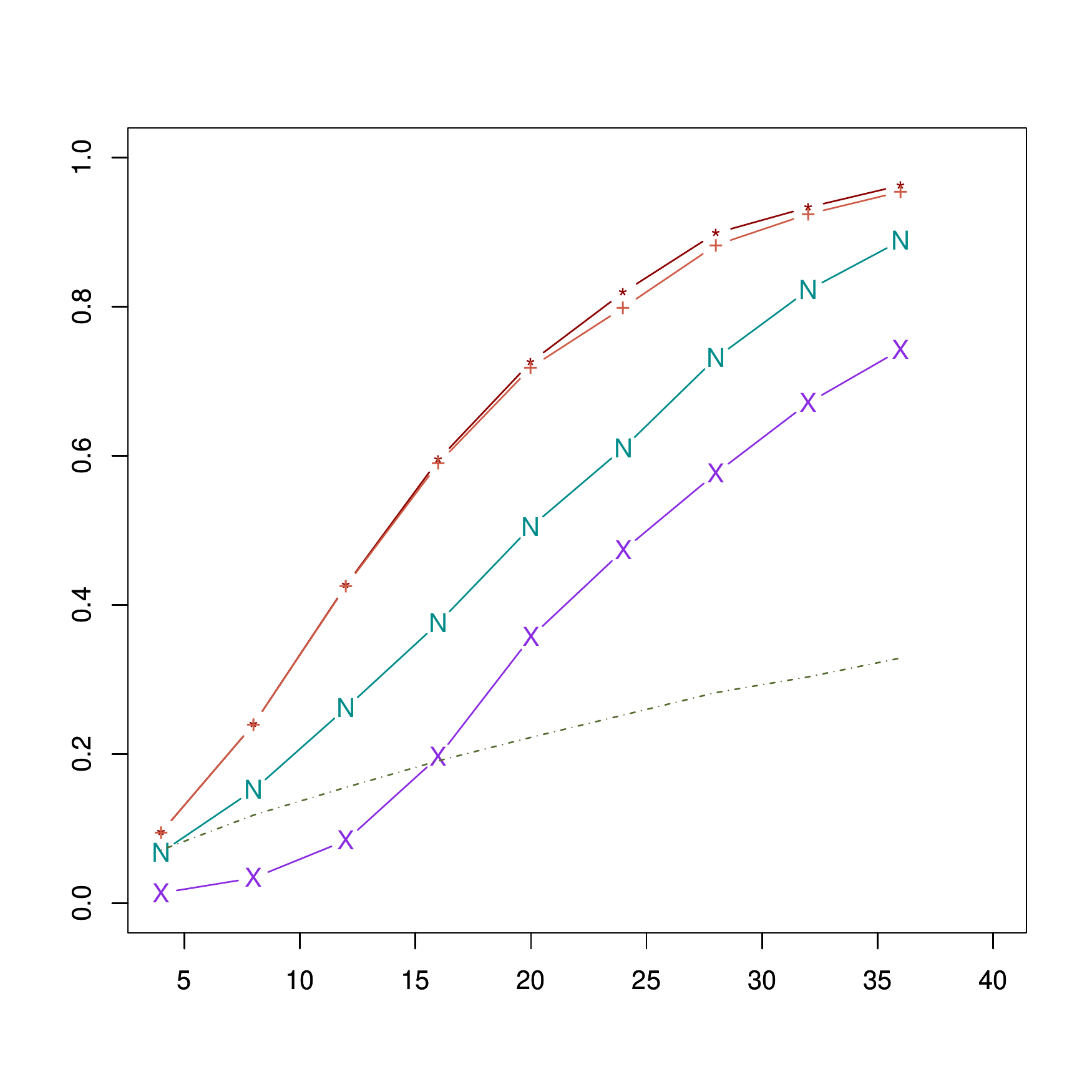} &
\includegraphics[width=0.333\textwidth]{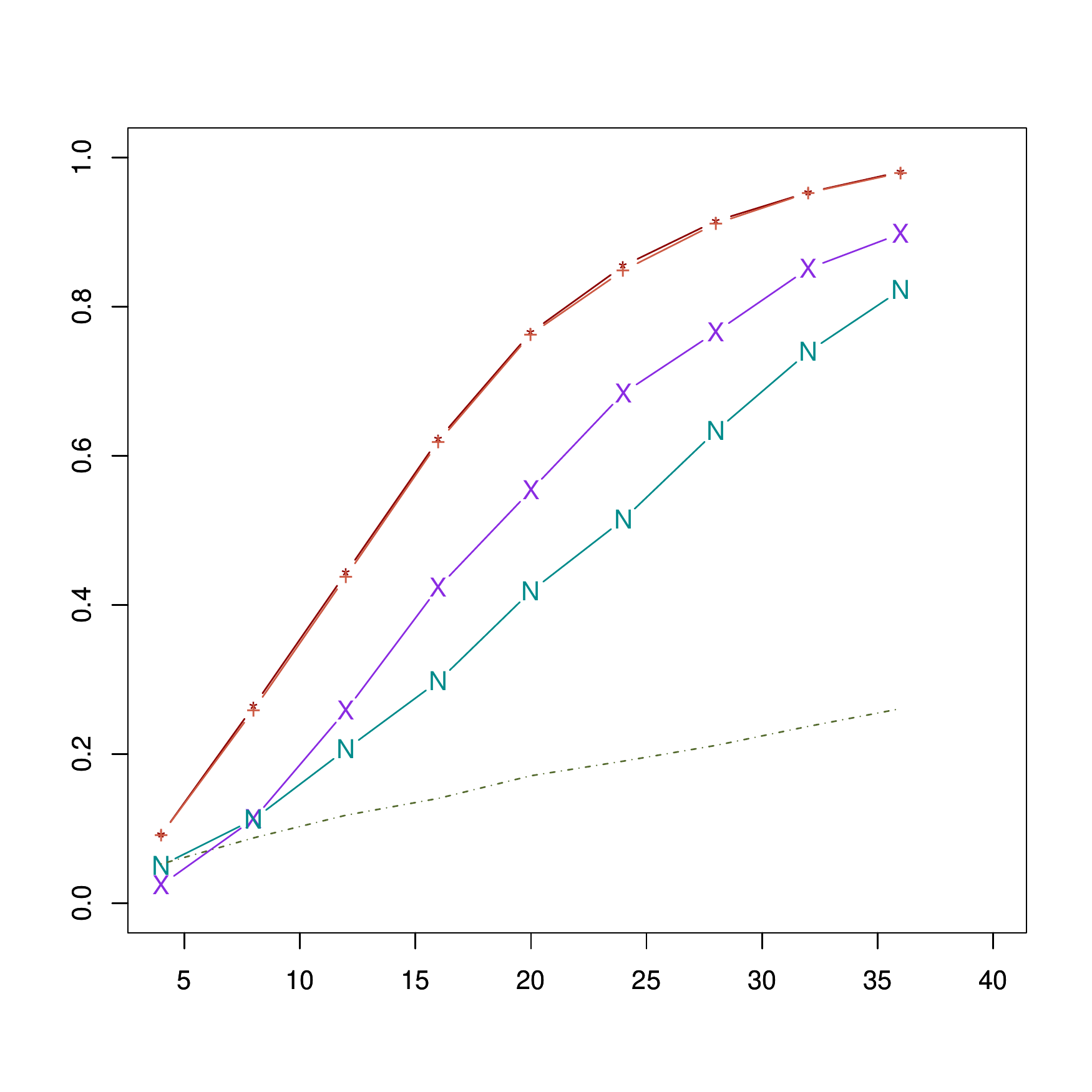} &
\includegraphics[width=0.333\textwidth]{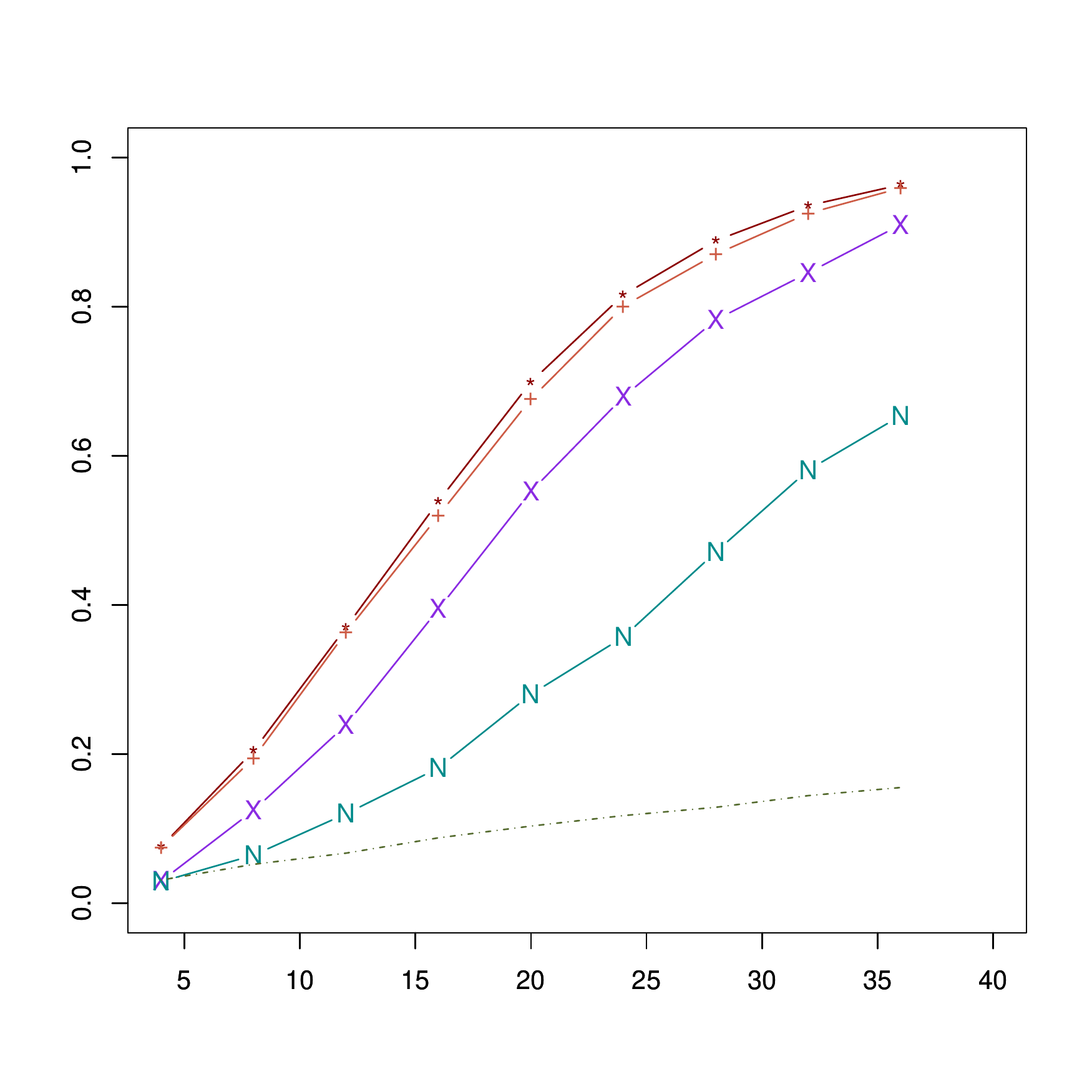} \\

\includegraphics[width=0.333\textwidth]{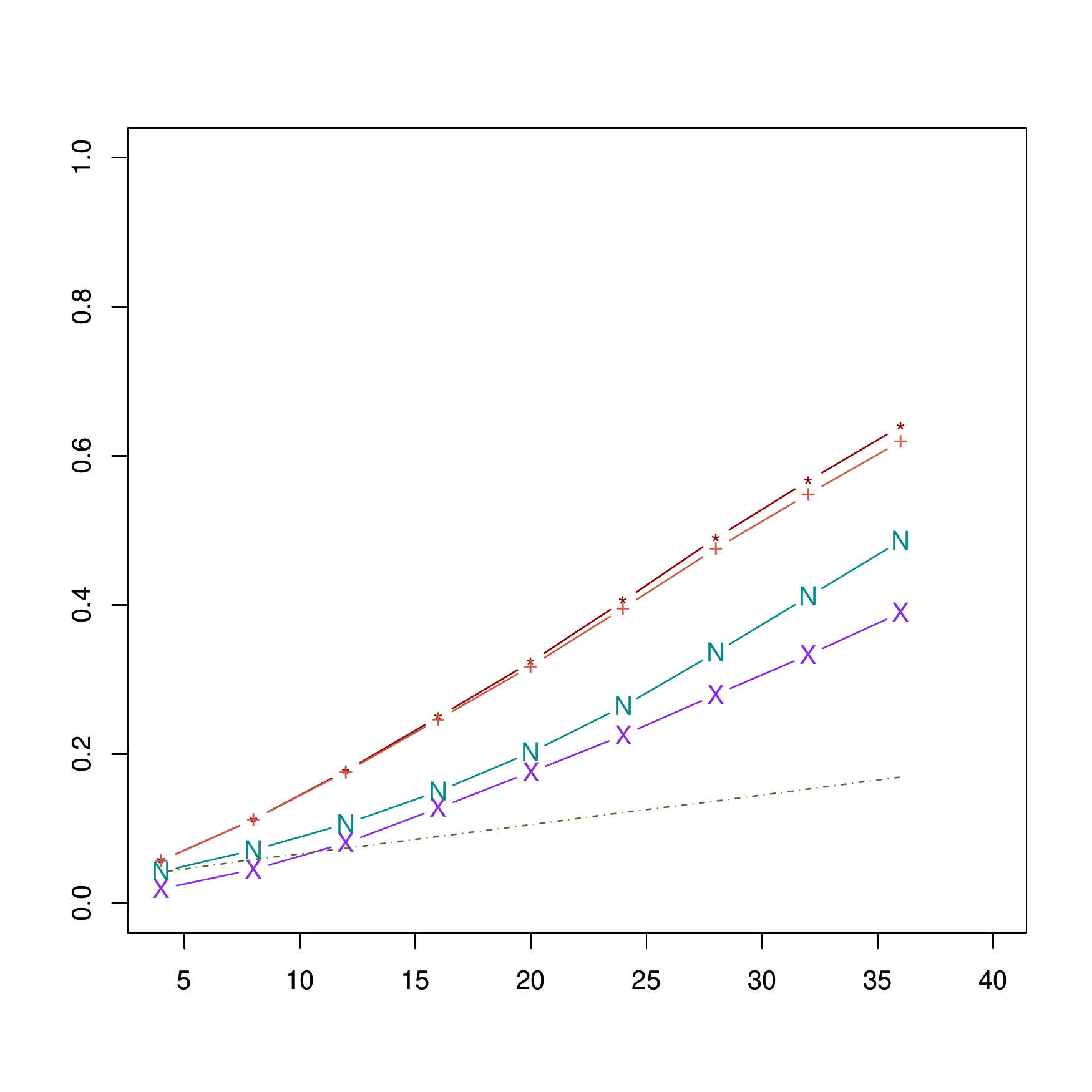} &
\includegraphics[width=0.333\textwidth]{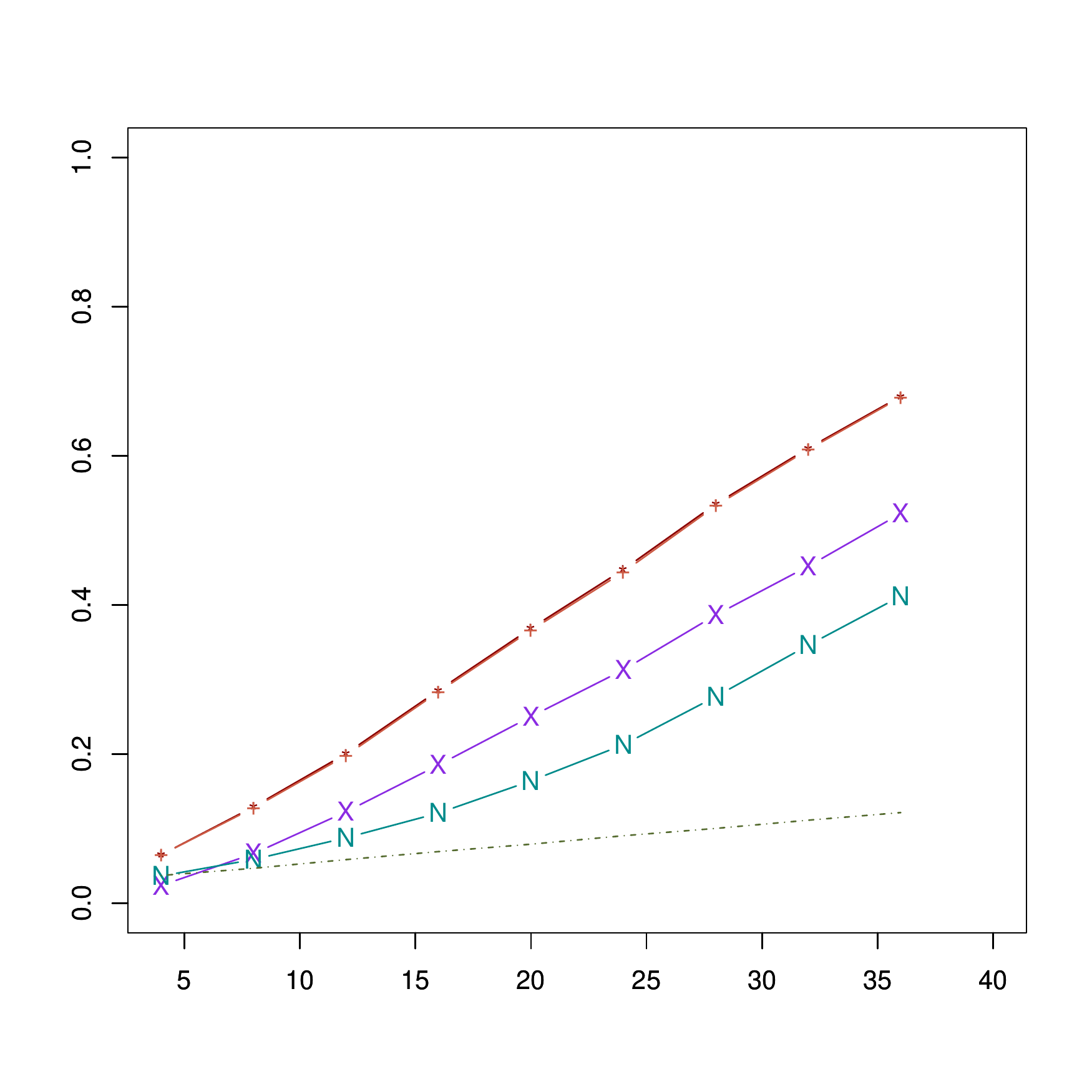} &
\includegraphics[width=0.333\textwidth]{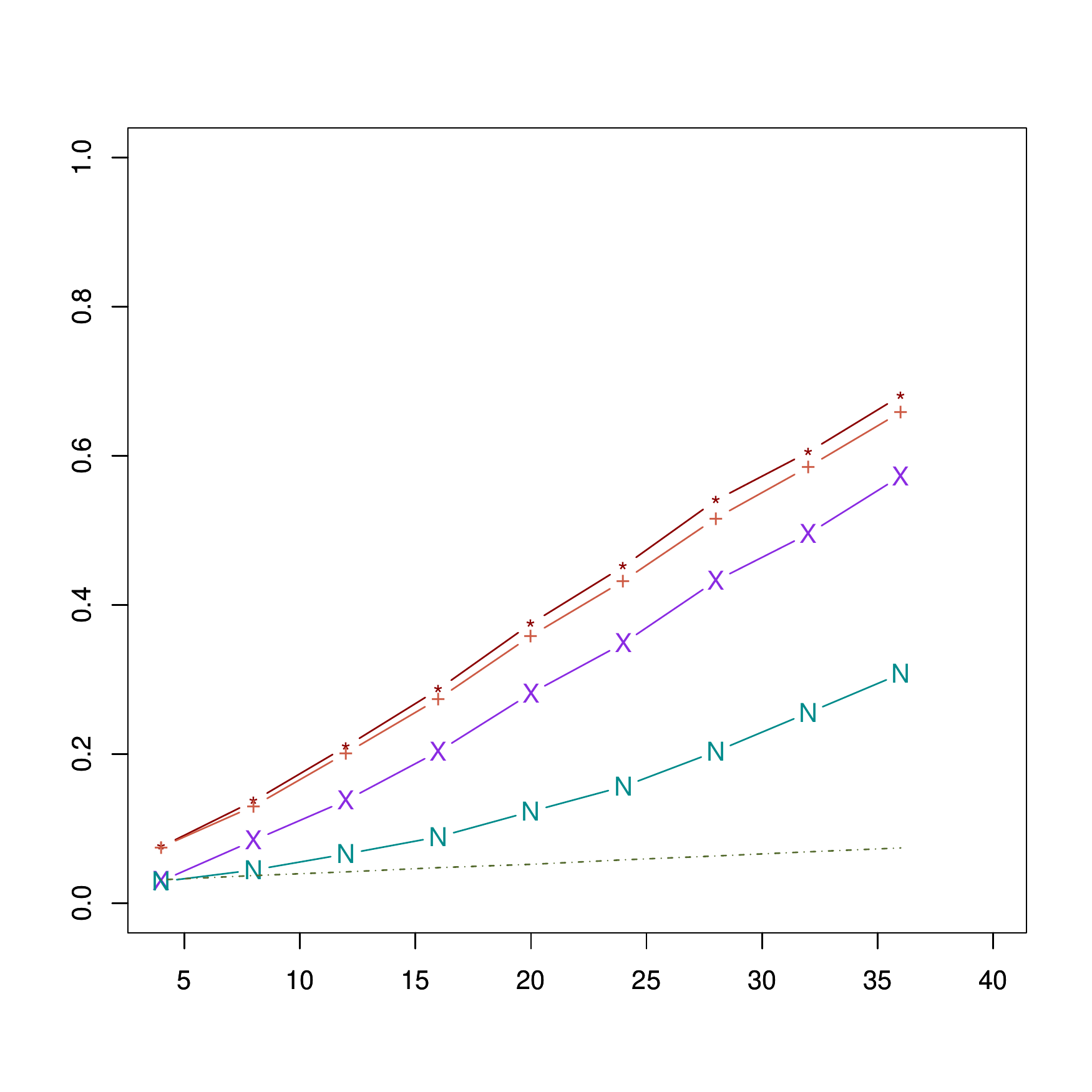} \\

\end{tabular}

\end{centering}

\caption{\label{sim_bymetric}
The performance of $\gamma = \{0,1,0.5,\gamma^\star\}$ according to
$E[\mbox{S@1}(\gamma)]$,
$\mbox{MRR}(\gamma)$, and
$\mbox{MAP}(\gamma)$
(top, middle, and bottom, respectively).
Columns, from left to right,
represent $m'=\frac{m}{4},\frac{m}{2},\frac{3m}{4}$.
The $x$-axis represents increasing values of $m$ and the $y$-axis represents the evaluation criterion.
Results are obtained via $1000$ random graphs generated according to
the $\kappa(n=184, p=[0.6,0.2,0.2]', m, s=[0.4,0.4,0.2]'; m')$ model.
As in the previous figure, lines with ``X'' markers denote context alone , those with ``N'' markers denote content alone,
those with ``+'' markers denote $\gamma=0.5$ and those with ``*'' markers denote $\gamma=\gamma^\star$.
The dashed line with no markers denotes chance performance.}

\end{figure}

Figure \ref{sim_bymetric}
generalizes the results presented in
Figure \ref{lg_sim_bymetric},
presenting performance
as we vary $m'$ (the proportion of $m$ with observed attributes)
for all three of our evaluation criteria.
Again, for small $m$,
all perform equally poorly (approximately chance).
As we vary the ratio of $m$ to $m'$, we see that
$T^{0.5}$ again is superior to either $T^1$ or $T^0$ alone in some cases
($m'=\frac{m}{4}, \frac{m}{2}$), but $T^0$ is superior to $T^1$ and $T^{0.5}$ in
other cases ($m'=\frac{3m}{4}$).
(Chance, indicated by the dashed green line, is not the same throughout
Figure \ref{sim_bymetric},
since we fix $n$ but vary $m'$:
the number of correct answers left in the candidate set \redcandidates, from left to right,
is $\frac{3m}{4}$, $\frac{m}{2}$, $\frac{m}{4}$. The performance of $T^1$ changes
across plots only because of these variations in chance performance.)

\begin{figure}
\begin{centering}

\includegraphics[width=400pt]{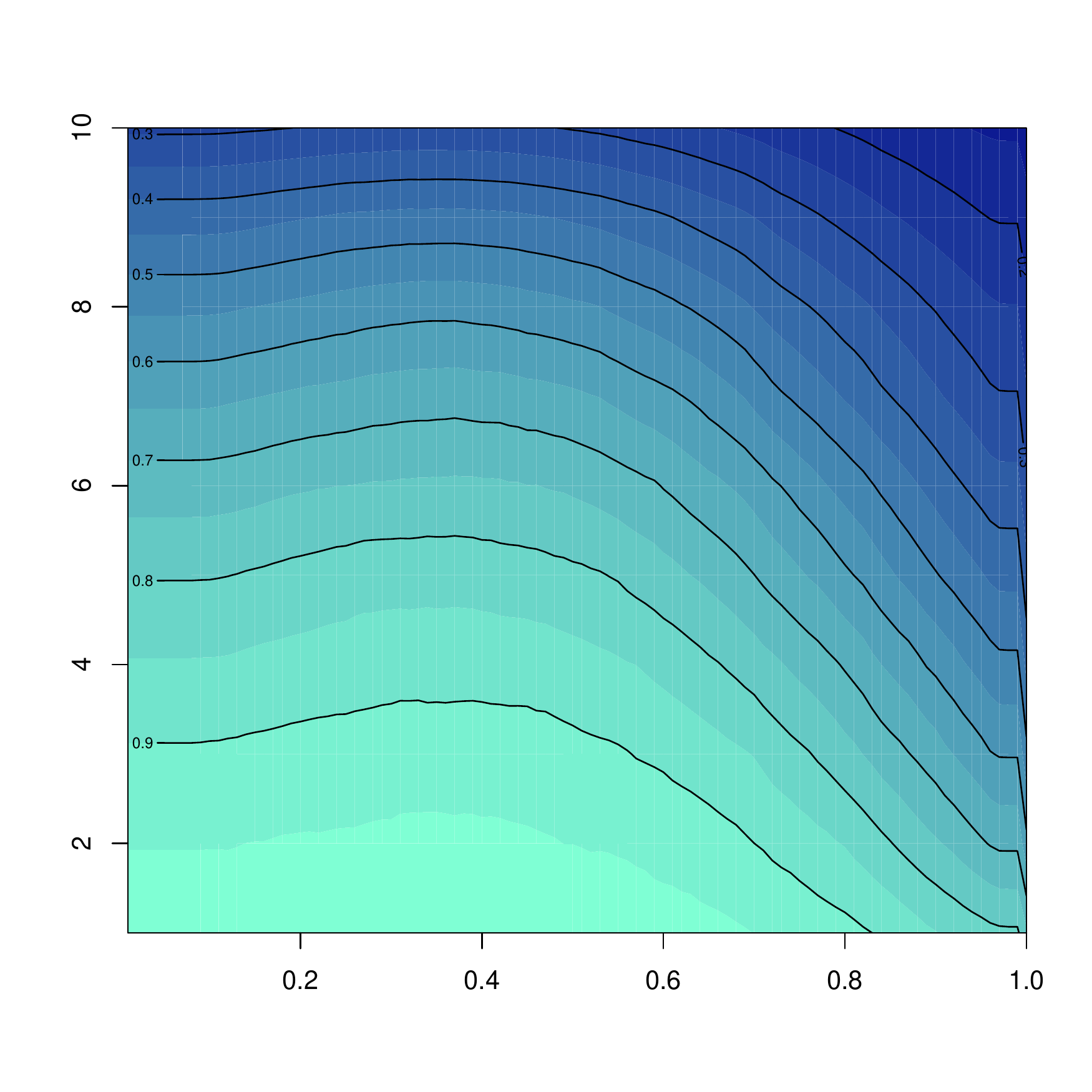}

\end{centering}

\caption{\label{sim_3d_roc}
The colors/contours of this plot denote $\mbox{AP}^y(\gamma)$, 
with the $y$-axis representing $y$ and 
the $x$-axis representing $\gamma$.
Note that $\gamma^\star$ (for all $y$ under consideration in this plot)
is found near $0.4$, as indicated by the increase in $\mbox{AP}^y(\gamma)$ in that region.  
Results are obtained via $1000$ Monte Carlo replicates.} 
\end{figure}

Figure \ref{sim_3d_roc} 
generalizes the results presented in
Figure \ref{lg_sim_bymetric},
by showing performance (measured in average precision) as a function of $\gamma$, 
with $\gamma$ free to vary from $[0,1]$. 
Let $k$ be the integer such that $V^\gamma_{(k)}$ is the $y^{\mbox{th}}$
highest ranked true but unknown red vertex, then 
\begin{equation}
\mbox{AP}^y(\gamma) = {\frac{\displaystyle\sum_{i=1}^{k}\mathbb{I}\{v^\gamma_{(i)} \in \redcandidates\}Pre(i,\gamma)}{y}}.
\end{equation}
For example, if we are to correctly identify $y=3$ true but unknown reds, then 
$\mbox{AP}^3(0.1) \cong 0.9$ and $\mbox{AP}^3(0.8) \cong 0.8$; 
For $y=5$, $\mbox{AP}^(0.1) \cong 0.8$ and $\mbox{AP}^3(0.8) \cong 0.7$
Observe that $\gamma^\star \in (0,1)$, rather than $\{0,1\}$,
indicating that the fusion of content and context can provide superior inferential power.


Thus we have demonstrated that fusion of content and context can be most effective, but is not always so.
We also have shown that the relative performance of content, context, and fusion depend upon $m$ and $m'$.
Further results, omitted for brevity, demonstrate that performance also depends on $n$, $p$ and $s$;
furthermore, even when fixing $p$, $s$, and $m'$,
there are scenarios where content is equal to, better than, and worse than context;
likewise, when $p$, $s$, and $m-m'$ are fixed.
So  the relative performance of content and context depends on more than the simple
relationship between $m$ and $m'$.
These relative performance phenomenon are present regardless of evaluation criteria.

\section{Experiments with Observed Graphs}
\label{observed_experiments}

We address three questions in this section:
(1) Do the phenomena described by our assumptions from Section \ref{the_model}
naturally occur?
(2) If and when these phenomena do occur, is the vertex nomination procedure laid out in Section \ref{methods} a viable approach to uncover occluded vertices?
(3) If and when these phenomena do occur and the vertex nomination procedure is viable, is it better to use context information alone ($\gamma=0$),
content information alone ($\gamma=1$) or a linear fusion of the two ($\gamma \in (0,1)$)?

Simulations provide useful insight into
how vertex nomination performs
when the phenomenon of interest is generated
according to a model based on our assumptions and
limited understanding of the underlying social
phenomena (Section \ref{simulations}).
Our simulations do not purport to capture all the salient
aspects of the human-generated behavior that gives rise
to the set of emails in our corpus. 
Thus, to investigate the efficacy of vertex nomination beyond
our generative model, we use importance sampling 
to discover naturally occurring examples of the phenomena of interest.
We then demonstrate that vertex nomination works for these naturally occurring phenomena.
We consider partitions of $V$ which satisfy our assumptions (from Section \ref{the_model}),
and estimate the parameters of a $\kappa$ graph model, for comparison to results from the generative model. 
Note that these are estimates of the parameter values from real data, rather than
set parameter values.

\subsection{Importance Sampling}
\label{importance_sampling}

We obtain a communications graph from the Enron email corpus;
$V$ is comprised of $n=184$ vertices (email addresses),
and edges connect pairs of vertices that communicate at least once during
a specific 20 week time period ($E_{Enron}$).
We consider Enron graph $G_{Enron}=(V,E_{Enron})$. (See Appendix for further details.) 

We augment $G_{Enron}$ with edge-topics 
in $\{1,\cdots,32\}$ 
obtained from \cite{BerryTopics}.
We fix $m=10$ and $m'=5$ for this section.
Given $m$, we randomly select a {\em candidate set} $\mM \subset V$.
We then evaluate the appropriateness of the disjoint partition $(\mM,\greenvertices)$
in terms of our two assumptions from Section \ref{the_model}:
 the first requires that the frequency of communications among pairs of vertices in $\mM$
 be higher than the frequency for other pairs, and
 the second requires a differential in topic distribution.
Toward this end, we consider for \textbf{Assumption 1}
\begin{equation}
  \Delta\rho = \rho(\Omega(\mM)) - \rho(\Omega(\greenvertices))
\end{equation}
where
$\Omega(V')$ is the subgraph in $G=(V,E)$ induced by the subset of vertices $V' \subset V$
and
the {\em relative density} of a graph $G=(V,E)$, $\rho(G)$, is defined as
$$\rho(G(V,E)) = \frac{|E|}{ {{|V|}\choose{2}} }.$$
For \textbf{Assumption 2}, we consider
\begin{equation}
  \Delta P = ||P(\Omega(\mM)) - P(\Omega(\greenvertices))||_1
\end{equation}
where the vector
$P(G) = [P_1(G), \cdots, P_{32}(G)]$ is the empirical distribution of edge-topics.
Thus $\Delta P$ represents the differential in topic distribution
between $\mM$ and $\greenvertices$.
In the Enron collection,
edges often represent multiple messages between the two email addresses, so for any edge $e$
we induce a probability distribution $\mathscr{T}_e$ over Berry topics $\{1, \cdots, 32\}$ 
from the observed messages;
note that for $e=uv$, $\mathscr{T}_e$ is just $P(\Omega(\{u,v\}))$.

Given the Enron graph $G_{Enron} = (V, E_{Enron})$ and specified $m$ and $m'$,
our importance sampling proceeds as follows:
\begin{enumerate}
\item Randomly partition the vertices into \mM\ and \greenvertices. 
\item If either $\Delta\rho \leq \tau_\rho$ or $\Delta P \leq \tau_{P}$ then discard this $(\mM, \greenvertices)$ partition and restart,
where $\tau_\rho$ and $\tau_{P}$ are somewhat arbitrarily specified thresholds.
\item Otherwise,
\subitem Label the vertices in $\mM$ $red$ ($\phi_V(v)=1$ for $v \in \mM$);
\subitem Label the vertices in
\greenvertices\ $green$ ($\phi_V(v)=2$ for $v \in \greenvertices$);
\subitem Define a mapping $\mathscr{M}$ from topic number $\{1, \cdots , 32\}$ to attribute $\{ red, green \}$
by letting $\Delta P_k = P_k(\Omega(\mM)) - P_k(\Omega(\greenvertices))$ for each topic $k$ and if $\Delta P_k > 0$, $\mathscr{M}(k) = 1$ ($red$); otherwise $\mathscr{M}(k) = 2$ ($green$).
\end{enumerate}

From this importance-sampling procedure, we have a set of acceptable vertex partitions $(\mM, \greenvertices)$ 
and corresponding mappings from Berry topics to $red$ or $green$ attributes ($\mathscr{M}$).
For each acceptable partition-map pair, we perform Monte Carlo experiments by instantiating each edge
with a single Berry-topic, according to its topic distribution, and proceeding with vertex nomination according
to the following procedure:

\begin{enumerate}
\item Draw a topic $T_e$ for edge $e$ according to its distribution over Berry topics $\mathscr{T}_e$.
\item Attribute each edge $e \in E$ with $\mathscr{M}(T_e)$ ($\phi_E(e)=\mathscr{M}(T_e)$).
\item Thus, $G(V,E,\phi_V, \phi_E)$.
\item Randomly select $\mM'$ from $\mM$ to be the vertices with observed vertex-attributes; occlude attributes on the rest of the vertices (\candidateset).
\item Thus, $G'(V,E,\phi_V,\phi_E,\mM')$.
\item Perform vertex nomination.
\end{enumerate}

From an observed graph obtained by the procedure described above (both importance sampling and instantiation),
we obtain estimates $\hat{{p}}$ and $\hat{{s}}$ by counting the proportion of
the possible edges that exist and have the appropriate attribute:
$$\hat{{p}}_1 = \frac{|\{e \in E(\Omega(\greenvertices)) : \phi_E(e) = 1\}|} { { {n-m}\choose{2} } },
  \hat{{p}}_2 = \frac{|\{e \in E(\Omega(\greenvertices)) : \phi_E(e) = 2\}|} { { {n-m}\choose{2} } }$$
  and
$$\hat{{s}}_1 = \frac{|\{e \in E(\Omega(\mM)): \phi_E(e)=1\}|} { { {m}\choose{2} } },
  \hat{{s}}_2 = \frac{|\{e \in E(\Omega(\mM)): \phi_E(e)=2\}|} { { {m}\choose{2} } }.$$
  For $\tau_P > 0$ and $\tau_\rho > 0$, this results in real Enron data attributed graphs
  satisfying (probabilistically) {\bf Assumptions 1} 
  and 
  {\bf 2}.


\subsection{Results for Enron Experiments}

\begin{figure}

\begin{centering}

\begin{tabular}{cc}

\includegraphics[width=0.5\textwidth]{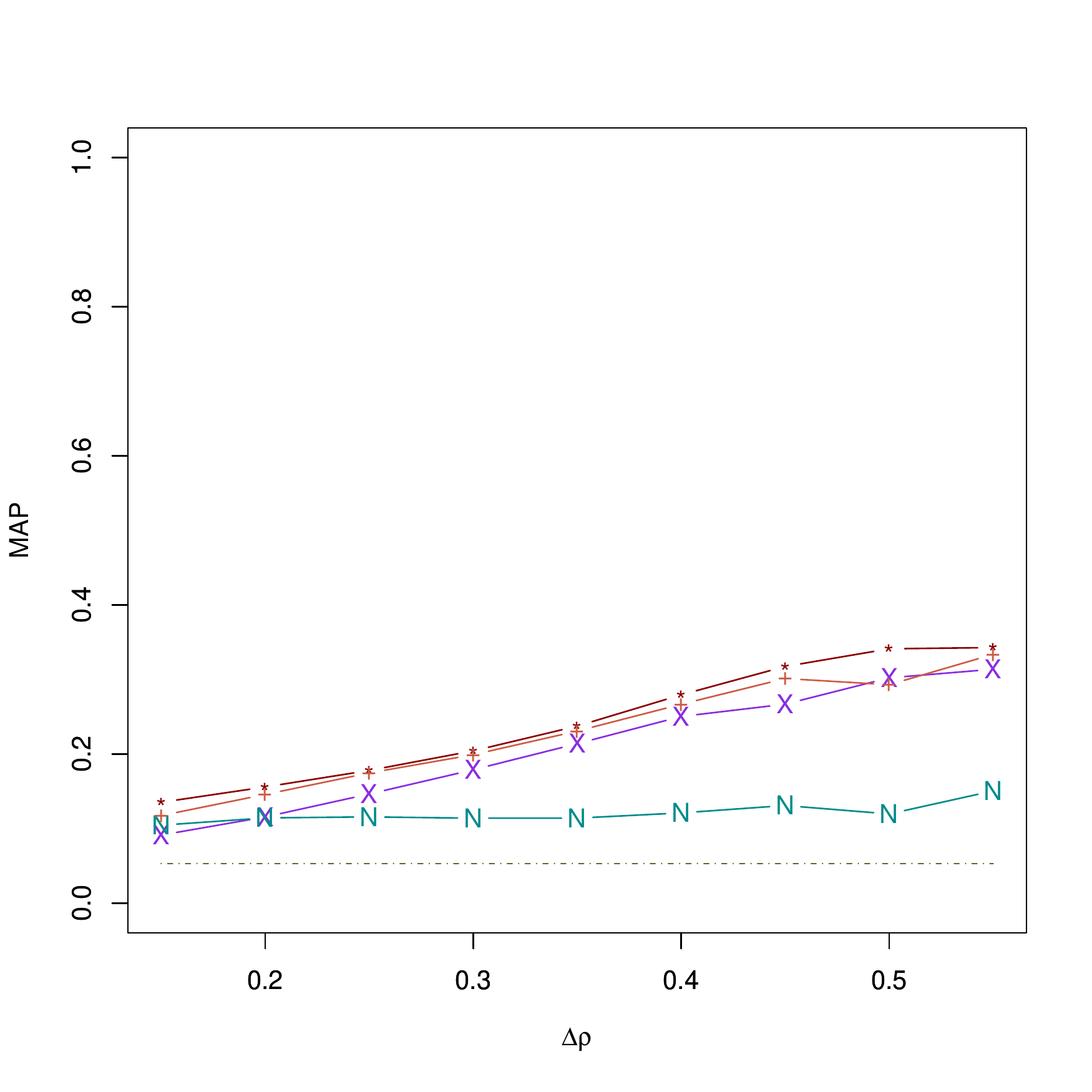} &
\includegraphics[width=0.5\textwidth]{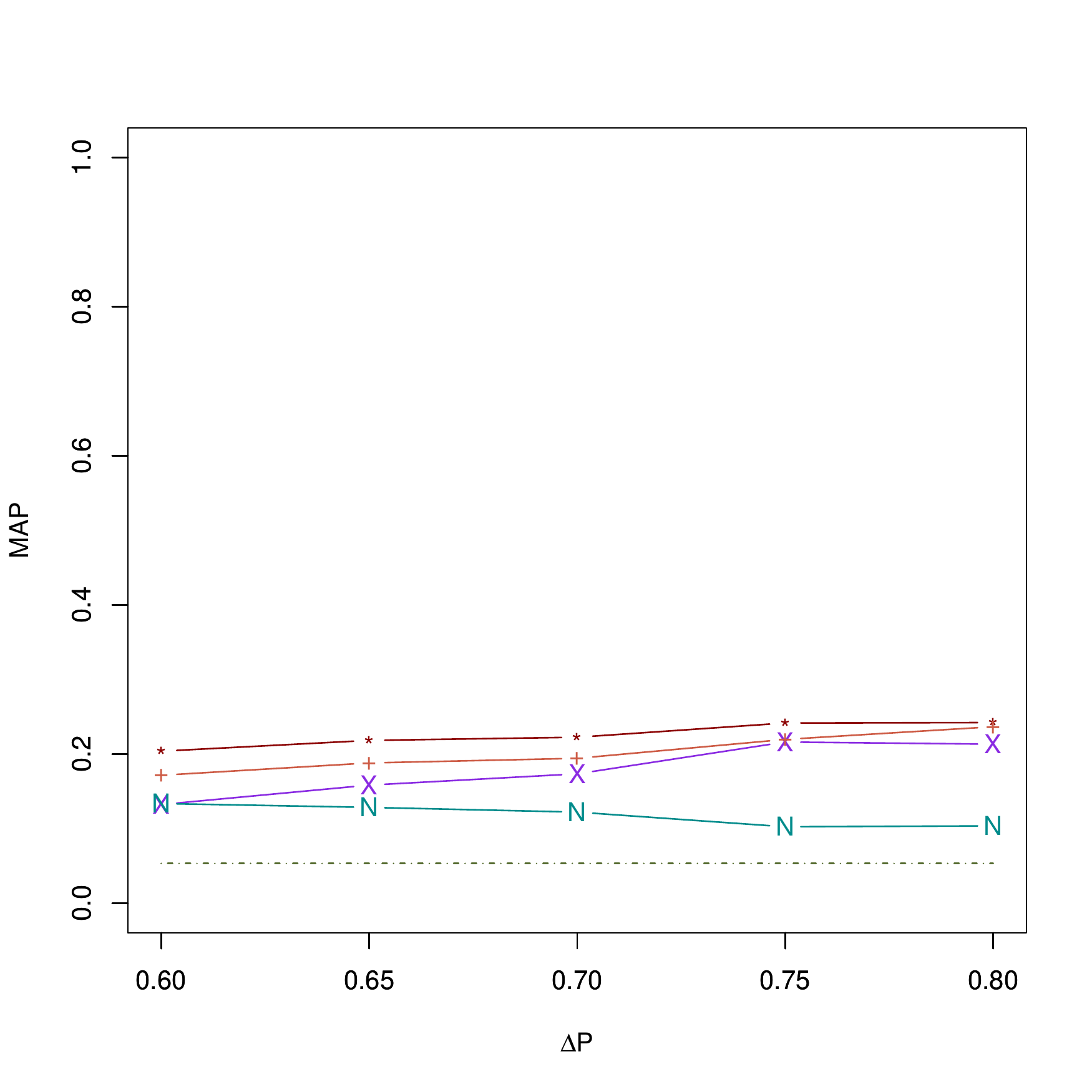} \\
	
\end{tabular}

\end{centering}

\caption{ \label{importance_conditioned_marginals} Content ('N'), context ('X'), arbitrary linear fusion ('+') and optimal linear fusion ('*') ($\gamma = \{0,1,0.5, \gamma^\star\}$ respectively) according to $\mbox{MAP}(\gamma)$ on importance-sampled graphs, plotted on the $y$-axis.
Left: $\mbox{MAP}$ ($y$-axis) and $\Delta\rho$ ($x$-axis), conditioned on a small range of $\Delta P \in [0.2,0.3]$.
Right: $\mbox{MAP}$ ($y$-axis) and $\Delta P$ ($x$-axis), conditioned on a small range of $\Delta \rho \in [0.3,0.4]$.
In all cases, the average reported reflects at least 20 partitions.
Chance is denoted by the dashed green line.}
\end{figure}

\begin{figure}

\begin{centering}

\includegraphics[width=0.55\textwidth]{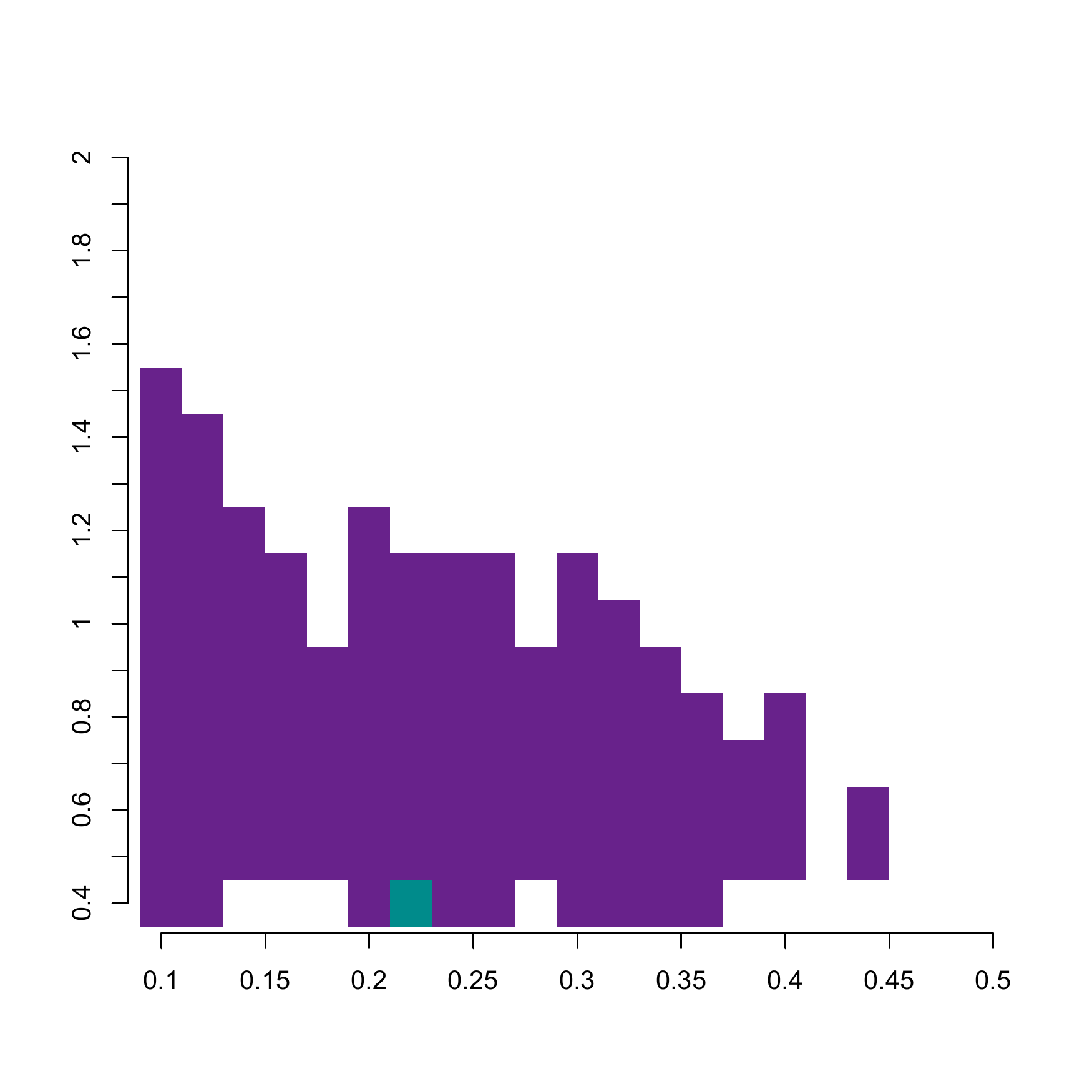}

\end{centering}

\caption{ \label{importance_joint_surface}
The difference in performance across
the joint space of $\Delta\rho$ and $\Delta P$ between additive fusion and content or context.
Specifically, we plot $\min(\mbox{MRR}(\gamma=0),\mbox{MRR}(\gamma=1)) - \mbox{MRR}(\gamma=0.5)$ for each $(\Delta\rho,\Delta P)$.
The $x$-axis shows $\Delta\rho$ and the $y$-axis shows $\Delta P$.
White indicates regions where there were an insufficient number of
observed samples to reliably calculate performance ($<20$).
Purple regions indicate that the performance at $\gamma=0.5 > \gamma \in \{0,1\}$ and
cyan regions indicate that performance at $\gamma=0 > \gamma \in \{0.5,1\}$.
Thus, purple indicates regions where fusion helps, cyan indicates regions where fusion hurts,
and white indicates regions where there is not enough data for a conclusive estimate.
White should also be interpreted as configurations that are highly unlikely, given the number of
samples investigated.
}
\end{figure}

Fusion of content and context generally yields an improvement
over either content  or context
alone, as shown in Figures \ref{importance_conditioned_marginals} and \ref{importance_joint_surface}.

\begin{figure}

\begin{centering}

\begin{tabular}{cc}
$\hat{p}_1$ & $\hat{p}_2$\\
\includegraphics[width=0.5\textwidth]{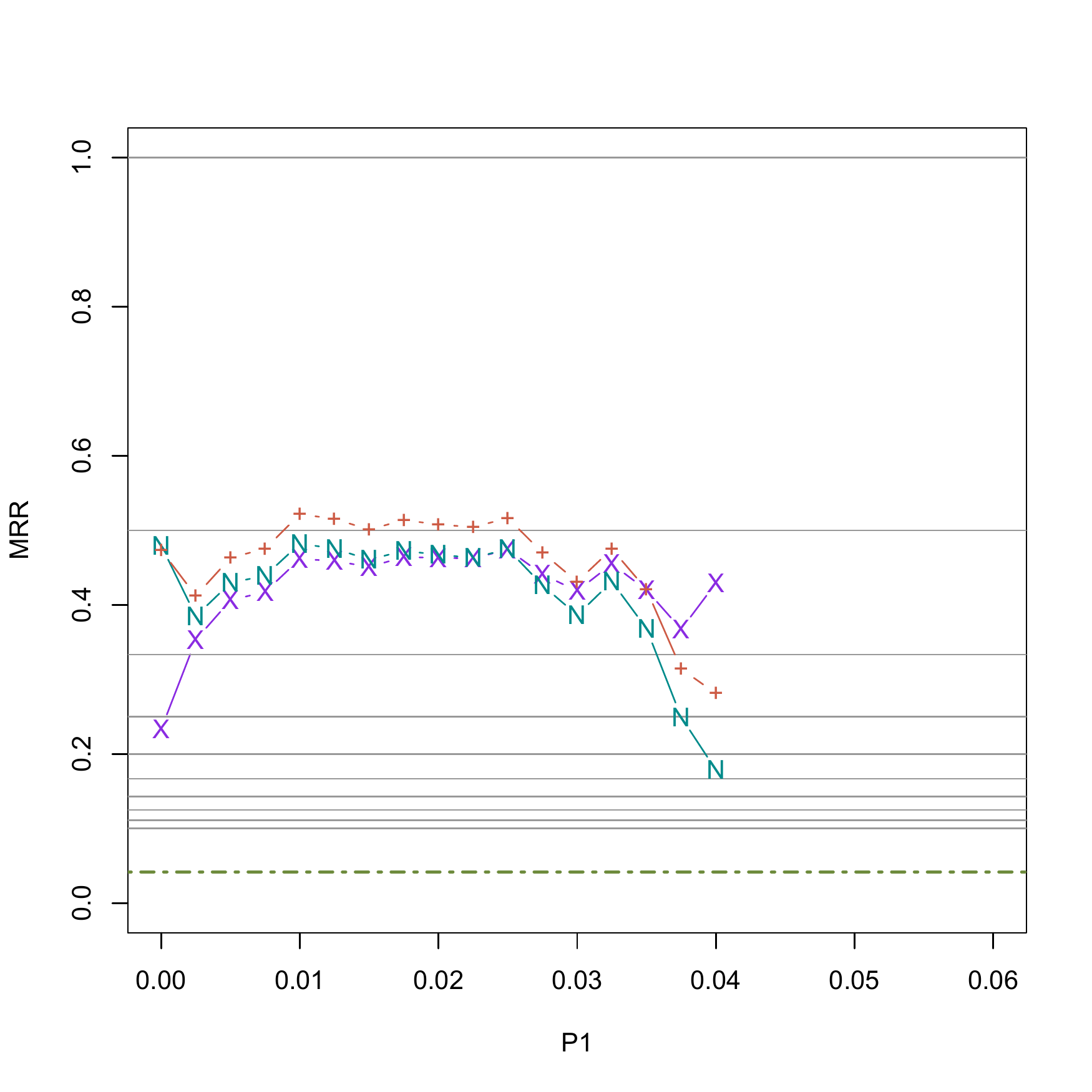} &
\includegraphics[width=0.5\textwidth]{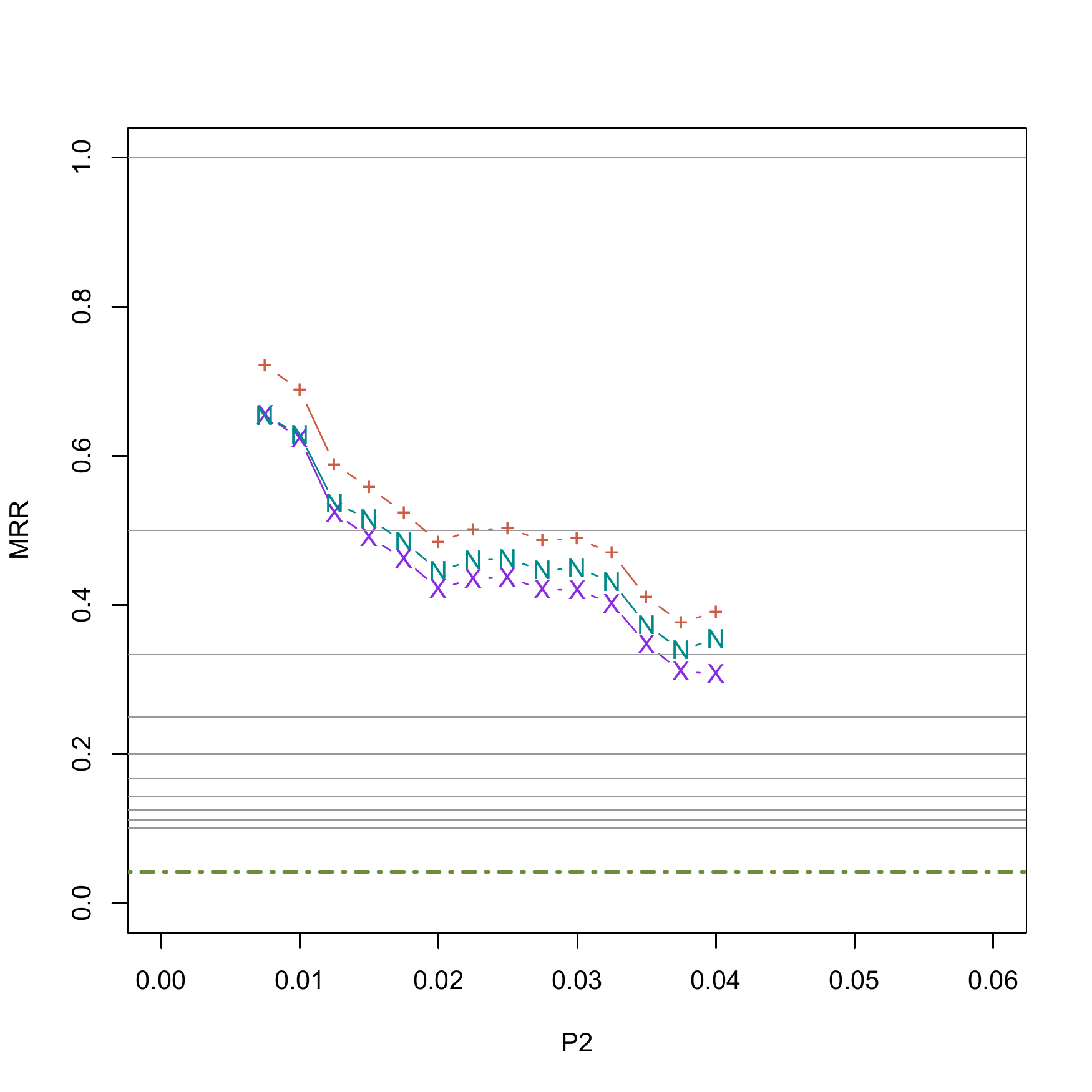} \\
\hline
$\hat{s}_1$ & $\hat{s}_2$\\
\includegraphics[width=0.5\textwidth]{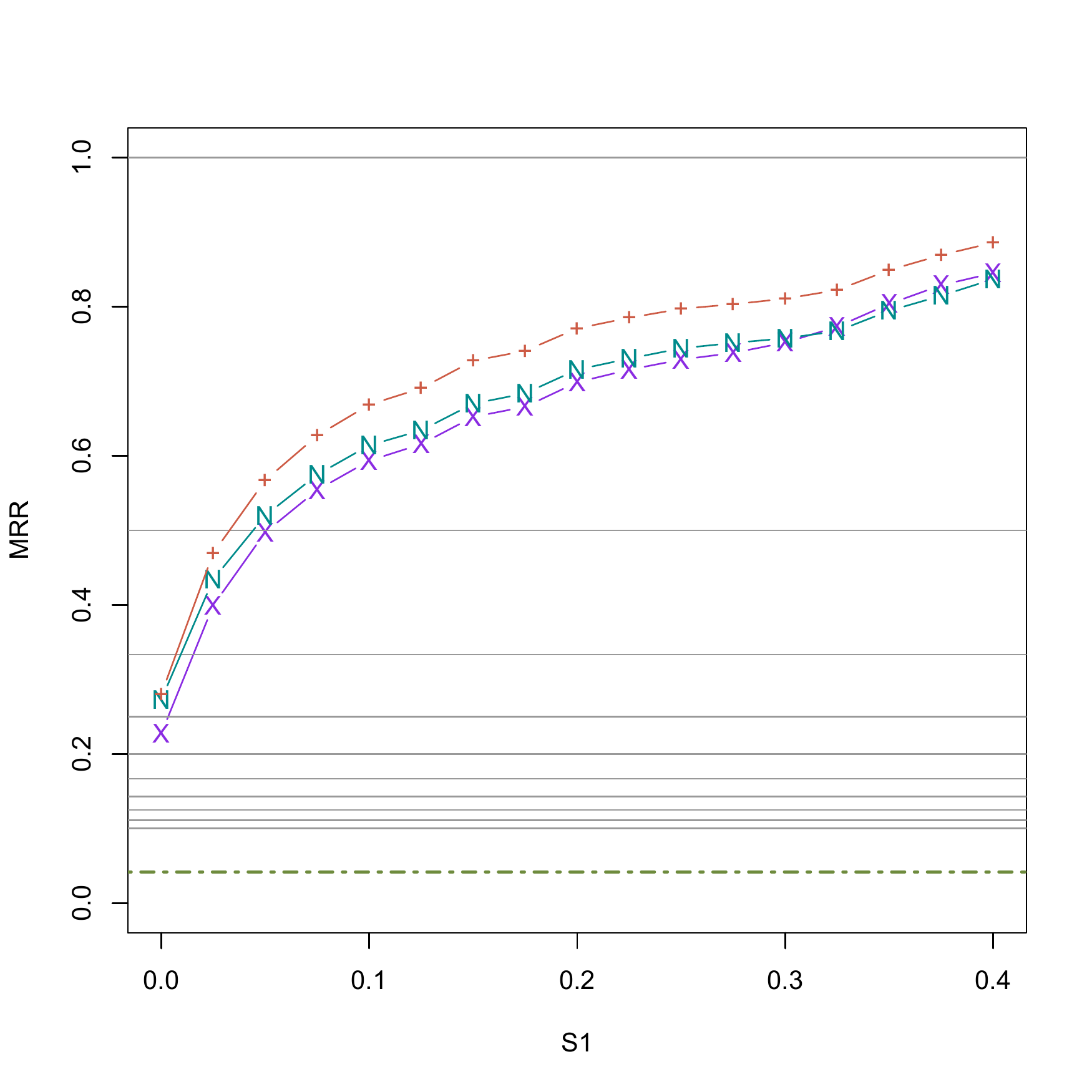} &
\includegraphics[width=0.5\textwidth]{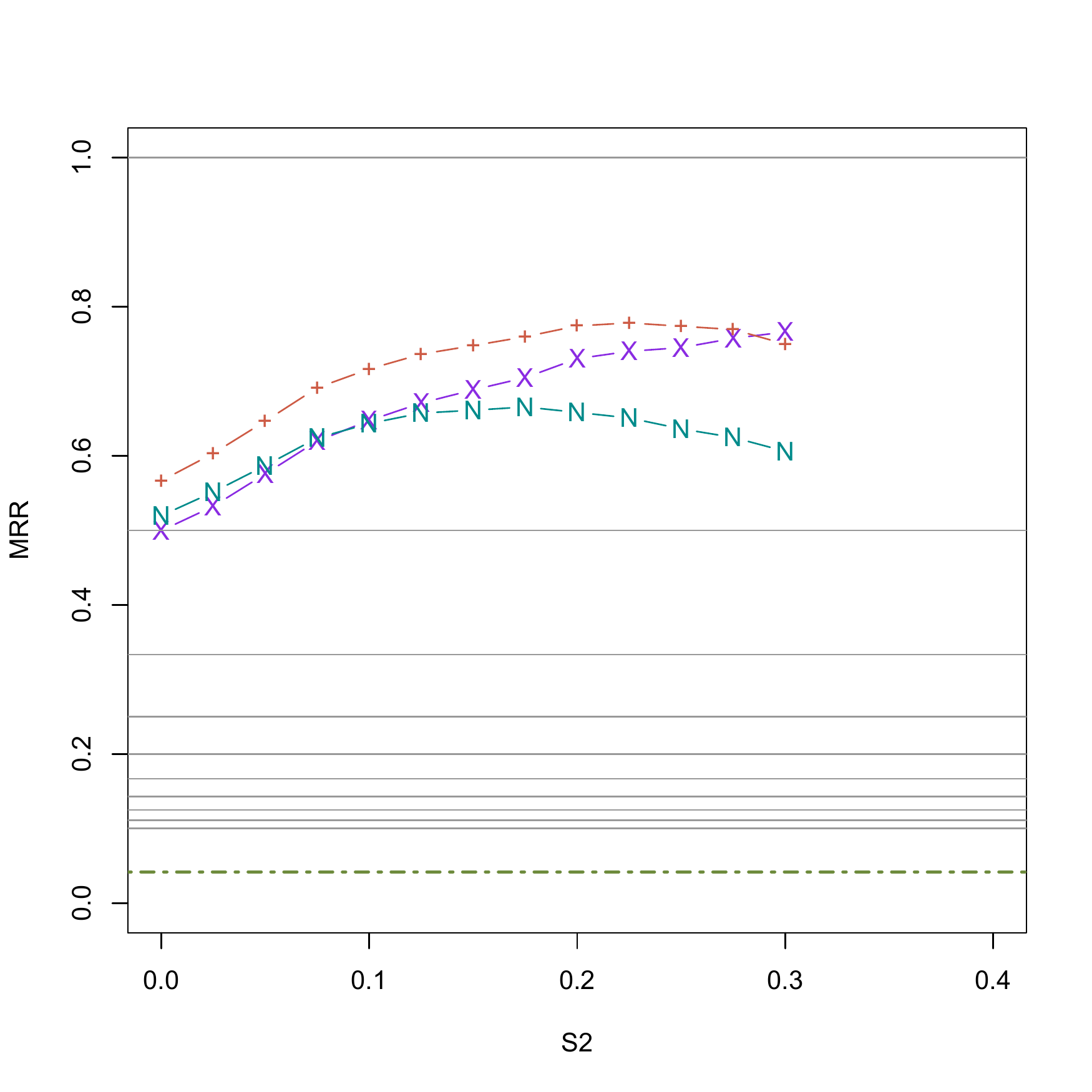} \\

\end{tabular}

\end{centering}

\caption{ \label{importance_RMinR_estimated_vectors} $\mbox{MRR}(\gamma)$ (on the $y$-axis) 
for $\gamma \in \{0,0.5,1\}$ 
for importance sampled graphs according to the estimated vectors $\hat{p}$ (top) and $\hat{s}$ (bottom). 
The $x$-axis denotes the estimated proportion of topic 1 (of interest) in the left column 
and topic 2 (not of interest) in the left column.
In all cases, the average reported reflects at least 20 partitions.
Chance is denoted by the dashed green line.
 }

\end{figure}

Figure \ref{importance_conditioned_marginals} reveals, as expected, that
the performance for $\gamma=0$ depends on $\Delta \rho$: as the $\mM$ are more
interconnected than are the $\greenvertices$, the probability of nominating a vertex
from $\redcandidates$ instead of $\greenvertices$ increases. Also, the performance for $\gamma=0$ is
largely independent of $\Delta P$.
Contrary to intuition,
perhaps, the performance of $\gamma=1$ is not wholly dependent upon $\Delta P$
nor entirely independent of $\Delta\rho$, due to the fact that the content signal 
depends on excess interesting content 
which is not independent of the probability of edges themselves.
Figure \ref{importance_RMinR_estimated_vectors} shows results comparable to plots
from other sections, as estimated from the importance-sampled observed graphs.


\begin{figure}
\begin{centering}
\includegraphics[width=400pt]{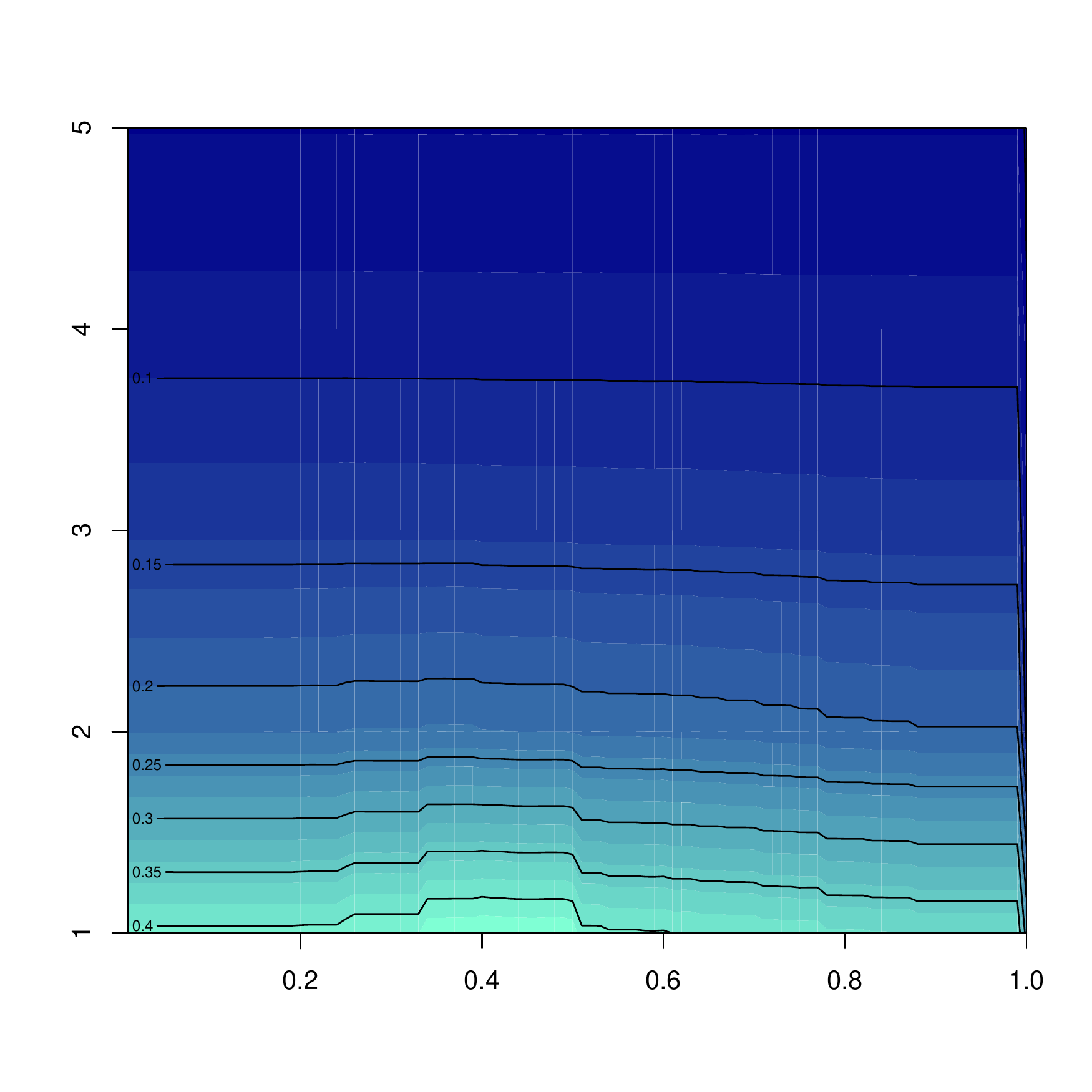}
\end{centering}

\caption{\label{is_3d_roc}
The colors/contours of this plot denote $\mbox{AP}^y(\gamma)$, 
with the $y$-axis representing $y$ and 
the $x$-axis representing $\gamma$.
Note that $\gamma^\star$ (for all $y$ under consideration in this plot)
is found near $0.4$, as indicated by the increase in $\mbox{AP}^y(\gamma)$ in that region.}
\end{figure}

Figure \ref{is_3d_roc} 
generalizes the results presented in
Figure \ref{importance_conditioned_marginals},
by showing performance (measured in average precision) as a function of $\gamma$, 
as $\gamma$ varies from $[0,1]$, for the importance sampled partitions present in a small
range of $\Delta \rho$ and $\Delta P$. 
Observe that $\gamma^\star$ is found for $\gamma \in (0,1)$, rather than $\gamma \in \{0,1\}$,
indicating that non-trivial fusion of content and context provides superior inferential power for this 
range of $\Delta \rho$ and $\Delta P$.

Figure \ref{importance_joint_surface} explores the differences between $T^{\gamma^\star}$
and $T^0$ or $T^1$ respectively, indicating where the performance
obtained by fusing content and context is greater than using either alone.
Where Figure \ref{is_3d_roc} reports results for a small range of $\Delta \rho$ and $\Delta P$, 
Figure \ref{importance_joint_surface} reports results for many such small regions (Figure \ref{is_3d_roc}
covers only one cell reported in Figure \ref{importance_joint_surface}). 
Over almost all the observed graphs obtained by the importance sampling
procedure, $T^{\gamma^\star}$ is superior to either $T^0$ or $T^1$.

In sum, we do find (1) that the phenomena of interest do naturally occur,
(2)  that when they do occur vertex nomination is viable, and 
(3) the fusion of content and context (arbitrary, $\gamma=0.5$ and optimal, $\gamma=\gamma^\star$) 
is superior to either alone for vertex nomination when these phenomena naturally occur.

\section{Conclusion}

\label{conclusions}

In this investigation we explore vertex nomination -- finding interesting vertices -- using information from context (graph structure) and content (edge-attributes). 
We present simulation and experimental results supporting
the intuition that content and context are often better together than either alone, for this task.

We present only simple linear content and context fusion statistics,
in an effort to demonstrate the fundamental superiority of non-trivial fusion. There is much
room for more complex and better performing content, context, and fusion statistics.
We leave this area
open to future research.

Results on real data are, by necessity, subjective for at least two reasons.
For one, the definition of ``interesting''
is likely to change significantly between datasets (and indeed those examining the datasets). 
Secondly, the relationship between the mathematical model ($\kappa$) and the 
observed behavior (and the effects thereof) is not easily quantified, so performance cannot
be easily predicted \textit{a priori}.
For illustrative purposes, we 
present results for one dataset (with one definition of interesting) using the Enron corpus,
though application and adaptation to new datasets (with different
interesting phenomena, behavior of vertices, and parameter values) remains an interesting and open question.

Knowledge of the relationships between performance and parameter values provides
useful information about the robustness and generalization of the techniques
beyond the simple setting explored.
Specifically, these relationships can be exploited when applying these (or similar)
techniques to real data, where analogs of the parameters can be estimated.

\newpage
\bibliography{bib}{}
\bibliographystyle{apalike}

\newpage
\section*{Appendix A: Enron Email Corpus}
\label{enron_email_corpus}

The Enron email corpus, used in this investigation, is a collection of emails
seized by the Securities and Exchange Commission (SEC) during their 
investigation into potentially fraudulent
and manipulative behavior of some Enron employees. 
Copies of all emails in the accounts of some 150 employees were obtained
(both send and received messages) and eventually released to the public.
We work with an approximately  27,000 message subset of the 500,000 
email messages seized (though some are duplicates found
in the inbox of many individuals), for comparison across studies (e.g. \cite{Priebe:2005}, \cite{mqEstimation})
The emails in the subset are those for which
both the sender and the receiver is on of a list of 184 employees
present in an organizational-heirarchy chart (mostly executives, traders, and secretaries). 
From this collection, we select an arbitrary 20 week period, 
from September 24, 2001 to February 11, 2002,
to examine.

Let $G_{Enron}=(V_{Enron},E_{Enron})$, where $v \in V_{Enron}$ is an email address
corresponding to one of the 184 employees mentioned above. $|V_{Enron}| = n_{Enron}$. If an email exists in the 
time period under consideration between $i,j \in V_{Enron}$, then $ij \in E_{Enron}$.
Approximately 5 percent of the ${n_{Enron}}\choose{2}$ possible edges exist, so  $\hat{p}_{Enron} = 0.05$.

A subset of emails (overlapping but different from that above) was labeled by topic by Michael Berry, 
\cite{BerryTopics}.
Bennett Landman, Tamer El-Sayed and Douglas Oard created a classifier, based on word count
histograms, for these Berry-topics. The entire dataset was labeled with this classifier (including
those originally labeled by Berry, for consistency). These topic labels are used throughout this investigation.

In fact, we treat each (undirected) edge $ij$ as the collection of messages exchanged between $i$ and $j$ during
the time period. This gives rise to our representation of each edge as a distribution over topics, since
each email has a topic associated with it, and each edge is comprised of many emails.

\end{document}